\documentclass[useAMS,usenatbib]{mn2e}
\usepackage{etex}
\usepackage{graphicx}
\usepackage{epstopdf}
\usepackage{stfloats}

\usepackage{epsf}
\usepackage{bm}
\usepackage{color}
\usepackage{times}
\usepackage{supertabular}
\usepackage{hyperref}
\usepackage{amsmath}
\usepackage{longtable}
\usepackage{tabularx}
\usepackage{booktabs}
\usepackage{array}
\usepackage{lscape}
\usepackage{amssymb}
\usepackage{threeparttable}
\usepackage{booktabs}
\usepackage{enumerate}
\title[Chemical abundance distribution function]{Metallicity distributions  of mono-age stellar populations of the  Galactic disc from the  LAMOST Galactic spectroscopic surveys}
\author[Wang et al.] {C. Wang$^{1}$\thanks{E-mails: wchun@pku.edu.cn (CW); x.liu@pku.edu.cn (XWL)},  
                 X.-W. Liu$^{1,2,3}$\footnotemark[1],  M.-S. Xiang$^{4,5}$\footnotemark[2], Y. Huang$^{1,3}$\footnotemark[2],B.-Q. Chen$^{3}$\thanks{LAMOST Fellow},  H.-B.Yuan$^{6}$,
                  \newauthor J.-J. Ren$^{4}$,  
                 H.-W. Zhang$^{1,2}$, Z.-J. Tian$^{1}$\footnotemark[2]\\ 
$^{1}$Department of Astronomy, Peking University, Beijing 100871, People's Republic of China.\\
$^{2}$Kavli Institute for Astronomy and Astrophysics, Peking University, Beijing 100871, People's Republic of China. \\
$^{3}$South-Western Institute for Astronomy Research, Yunnan University, Kunming, Yunnan 650091, People's Republic of China.\\
$^{4}$National Astronomical Observatories, Chinese Academy of Sciences, Beijing 100012, People's Republic of China.\\
$^{5}$ Max-Planck Institute for Astronomy, Konigstuhl, D-69117, Heidelberg, Germany. \\
$^{6}$Department of Astronomy, Beijing Normal University, Beijing 100875, People's Republic of China.}
\begin{document}
\date{20180530}
\pagerange{\pageref{firstpage}--\pageref{lastpage}} \pubyear{except for 2018}
\maketitle

\begin{abstract}
We have investigated the metallicity distributions of mono-age 
stellar populations across the disc of $6 \lesssim R \lesssim 12$\,kpc and $|Z| \lesssim 2$\,kpc 
using    samples selected from the main-sequence turn-off and sub-giant (MSTO-SG) stars 
targeted by the LAMOST Galactic Spectroscopic surveys.
Both the mean values and the profiles of the distributions exhibit
significant variations with age and  position. We confirm that the oldest ($>11$\,Gyr) 
stars have nearly flat  radial  [Fe/H] gradients at all heights above the disc but show negative vertical [Fe/H] 
gradients. For stars younger than 11\,Gyr, the radial  [Fe/H]  gradients flatten 
with $|Z|$, while the vertical  [Fe/H] gradients flatten with $R$. Stars of 4--6\,Gyr exhibit steeper 
negative radial  [Fe/H] gradients than those of either younger or older ages. 
Values of [$\alpha$/Fe] of mono-age stellar populations also show significant radial and vertical gradients, 
with patterns varying with age. The [Fe/H] distribution profiles of  old ($>8$\,Gyr) stars vary little with $R$,       
while those of younger stars exhibit strong radial variations, probably a consequence of 
significant radial migration. The  [$\alpha$/Fe] radial distribution profiles show opposite 
patterns of variations with age compared  to those of  [Fe/H].  
We have also explored the impacts of  stellar mixing
by epicycle motions (blurring) on the [Fe/H] and [$\alpha$/Fe] distributions, and found that  blurring  mainly change the widths of the distribution profiles. 
Our results suggest that the disc may have experienced a complex assemblage history, in which both 
the ``inside-out'' and ``upside-down'' formation processes may have played an  important role. 

\end{abstract}


\section{INTRODUCTION}

As a typical disc galaxy,  our own galaxy the Milky Way (MW) is an excellent laboratory to test the galaxy formation and  evolution scenarios with the advantage that the individual  stars in the MW can be resolved. Since most of the MW's baryons and angular momentum are  in the disc,  the knowledge of how the disc was  formed  and assembled is a key to understanding  the formation and evolution of our Galaxy.  However,  many questions regarding the disc   formation  and assemblage  remain open, and there is great debate about this topic \citep[e.g. ][]{Rix_review}. 

The surface chemical composition of  low- and intermediate-mass stars remains almost unchanged during the   main-sequence and turn-off evolutionary stage, which makes it fossil record  of the environment of the interstellar medium at the position and time of  the births of those stars. The stellar number density $N$ as a function of metallicity [Fe/H] and $\alpha$-element to iron abundance ratio [$\alpha$/Fe] at a given disc position of radius $R$ and height $Z$ and age $\tau$, the so-called  stellar metallicity distribution function, MDF $\equiv$ $N ( $[Fe/H], [$\alpha$/Fe] $|$ $R$, $Z$,  $\tau)$, is  the consequence of  many complex  physical processes, including the infall and outflow of  gas \citep{Larson,Colavitti2008,Pezzulli2016, Andrews2017,Toyouchi2018, Grisoni2018}, the supernova feedback \citep{Kobayashi}, the secular evolution \citep[e.g.  radial migration;][]{sellwood, Roskar2008, Schonrich, Loebman2011, Haywood2016}, the non-axisymmetric perturbations (e.g.  by the central bar or spiral arms), and  the accretions of dwarf galaxies \citep{Quinn1993}.  The stellar  metallicity distributions  therefore  provide strong constraints on the Galactic disc formation and evolution history, especially on  when, where and how much those processes affect the assemblage and evolution of the Galactic disc.   

The gradients of [Fe/H] and  [$\alpha$/Fe]  represent  the  first-order  variations  with spatial position (the slopes) of the zero-order MDF (the mean values).  The   [Fe/H]  gradients  in the radial and vertical directions have been determined  with a variety of tracers, including  OB stars  \citep{daflon}, Cepheid variables  \citep{Andrievsky, Luck}, H~{\sc ii} regions \citep{Balser},  open clusters  \citep{Chen,Magrini}, planetary nebulae  \citep{Costa,Henry} , FGK dwarfs  \citep{Katz,Cheng,Boeche}, (red) giant stars \citep{Hayden_2014,Boeche2014} and  red clump stars  \citep{Huang}.  The results reveal that the disc has  negative  gradients in both radial  and vertical  directions. The gradients have also been  demonstrated to exhibit  significant spatial variations.  The observed negative radial [Fe/H] gradients generally  support the ``inside-out'' disc forming scenario. 
On the other hand, none of the above studies clearly show  how the  [Fe/H] gradients vary among different stellar populations due to the lack of robust age  estimates for a large sample of stars. 
In a pioneering study,  \cite{Xiangmdf}  determine the radial and vertical [Fe/H] gradients of  mono-age stellar populations using a sample of 0.3 million   main-sequence turn-off (MSTO) stars with robust age estimates selected from the LAMOST Spectroscopic Surveys of the Galactic Anti-centre \citep[LSS-GAC;][]{liu-lss-gac,yuan-lamost}.  
They find that both  the radial and vertical gradients show  strong variations among different mono-age stellar populations. 
The  radial  gradients are very  small for the oldest populations, and   become steeper as age decreases, reaching a maximum (steepest) at age $6-8$ Gyr, and then  flatten again with  decreasing age. The vertical  gradients     generally flatten   with decreasing age   for $\tau < 11$ Gyr.       Their results   suggest that  the disc may  have experienced   at least two different assemblage phases. 

The MDF profiles  contain rich  information of the Galactic disc assemblage history.   The profile of the   [Fe/H]  distribution     in the solar neighbourhood has been  widely studied \citep[e.g.  ][]{van, Nordstrom,  Casagrande}.   The results show  very broad [Fe/H] profiles at the solar neighbourhood, suggesting  that both gas dynamics and stellar migration have played   important roles  in the disc assemblage history. 
Beyond  the solar neighbourhood, the  variations of the [Fe/H]  distribution   with  positions have been investigated  with data from the  RAVE \citep{siebert}, SEGUE \citep{schlesinger} and  APOGEE   \citep{alpha,Hayden_2015} surveys.    The results show that the profile of  the [Fe/H]  distribution varies with both $R$ and $Z$.  \cite{Hayden_2015}   demonstrate  that  the low-[$\alpha$/Fe] populations close to the Galactic plane have   negatively skewed [Fe/H] distributions in the inner disc and  positively skewed distributions in the outer disc, a result that  they interpret  as a  consequence of  stellar radial migration, in particular that induced by  the churning process. Their results also show that     high-[$\alpha$/Fe]  populations exhibit a similar profile  at all positions. The different behaviours of the MDF between the low-[$\alpha$/Fe]   and high-[$\alpha$/Fe]  populations suggest that  the profile of [Fe/H] distribution must vary with  population since     [$\alpha$/Fe] is   a  rough  indicator of stellar  age.   [$\alpha$/Fe]  however can  only  be used to differentiate stars  into two  populations: a young and  an old population, insufficient  to study in detail how the  strength of radial migration varies with time throughout the whole assemblage history of the disc.     A large sample of stars with  robust age estimates  is essential to clarify   this important  issue. 

Recently, robust ages  for  a million main-sequence turn-off and sub-giant stars (MSTO-SG),   selected  from the LAMOST  Galactic spectroscopic surveys \citep{Cui,lamost,deng-legue,liu-lss-gac,yuan-lamost}, have become available \citep{Xiang_msto}.  
The sample stars  span a wide range of age, cover  a large and  contiguous volume of the Galactic disc and have a simple yet non-trivial uniform target selection function.  
The sample   provides an excellent opportunity to  characterise   the   MDF  of  mono-age  stellar populations, and  understand  the disc assemblage  history in more detail.  In this work,  we present the MDF  of mono-age stellar populations across the disc of 6\,$\lesssim R \lesssim$\,12\,kpc and $ |Z|\lesssim$\,2\,kpc.  We have derived the distribution function  not only of [Fe/H], but also  of [$\alpha$/Fe]. Although there have been  some work focusing on the distributions of [$\alpha$/H] \citep{Genovali2015, Hayden_2015} and [$\alpha$/Fe] \citep{alpha,Boeche2014},  the study is not as comprehensive as that of [Fe/H]. 
 In the current work, we  investigate the spatial variations of both the gradients   and  profiles of the [Fe/H] and [$\alpha$/Fe] distributions for different mono-age stellar populations, and  explore the impacts  of stellar mixing by epicycle motions (blurring) on the distributions.  
Amongst them,  the radial and vertical [$\alpha$/Fe] gradients and the profiles of [Fe/H] and [$\alpha$/Fe] distributions  of mono-age stellar populations  are investigated  for the first time.

 \begin{figure*}
 \centering
 \includegraphics[width=6.5in]{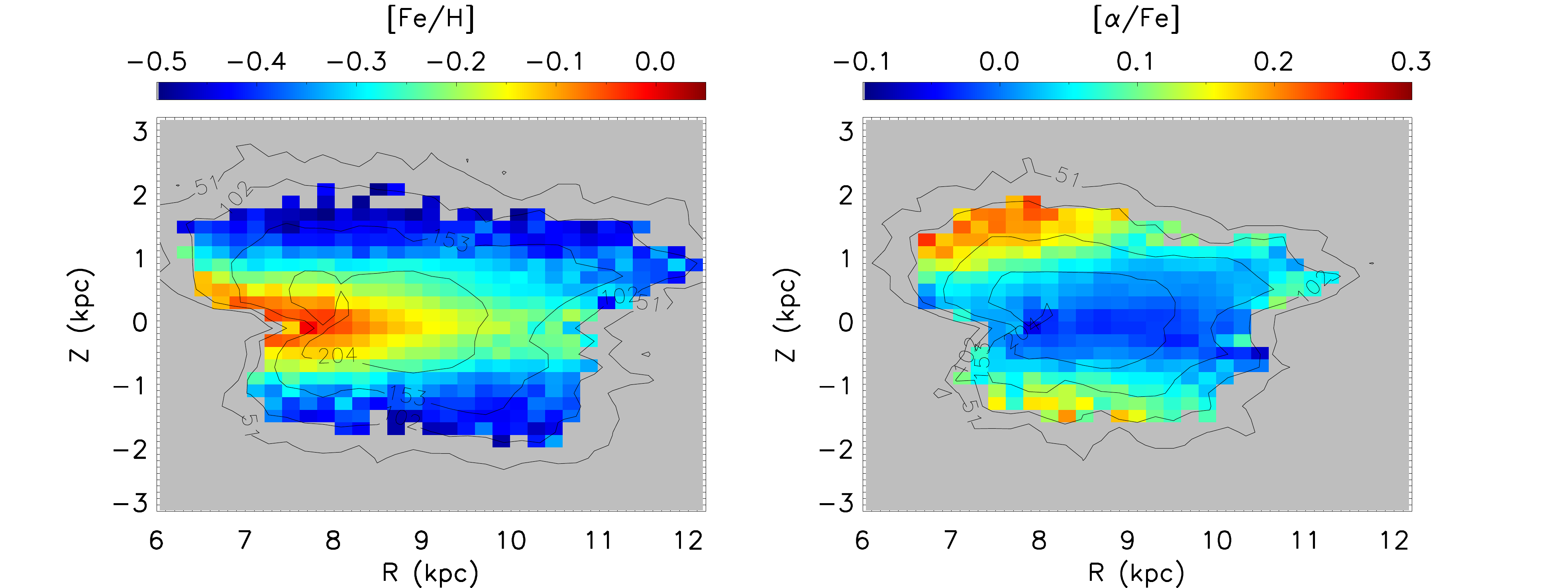}
 \caption{[Fe/H] (left panel) and [$\alpha$/Fe] (right panel) distributions of the MSTO-SG-SNR20 (left panel) and  MSTO-SG-SNR50 (right panel) samples,  binned by 0.2\,$\times{0.2}$\,kpc in  the $R$--$Z$ plane.  The over-plotted contours show the corresponding density (star number count  in each  $R$--$Z$ bin) distributions  on  logarithmic scale. }
\label{spatial_distribution}
 \end{figure*}
 
This paper is  organised as follows.   In Section 2,  we briefly introduce the MSTO-SG   sample and  corrections for the selection effects.  In Section 3, we  present the [Fe/H] and [$\alpha$/Fe] gradients of  mono-age stellar populations.  We present the metallicity distribution  profiles of mono-age stellar populations in Section 4. The effects of the  blurring process on the   metallicity distributions are discussed  in Section 5. The  constraints of the MDF derived here on the   formation and evolution of the  Galactic disc  are discussed in Section 6.  Finally, we summary our work in Section\, 7. 
\section{Data}
\subsection{The LAMOST MSTO-SG star sample}
The sample used in the current work consists of 0.93 million MSTO-SG  stars with robust age estimates selected from  the Pilot Surveys and the first four years of five-year Phase I  Regular Surveys of the LAMOST Galactic spectroscopic surveys \citep{Xiang_msto}.  
Several studies have been devoted to derive robust estimates of  stellar atmospheric parameters from the LAMOST spectra, such as the LAMOST Stellar Parameter Pipeline \citep[LASP;][]{lamost-dr1},  the LAMOST Stellar Parameter Pipeline at Peking University \citep[LSP3;][]{lsp3,Xiang-lss-gac}, applying  the SEGUE Stellar Parameter Pipeline (SSPP) to LAMOST spectra \citep{Lee2015},  parameter determinations based on the $Cannon$ \citep{Ho2017} and an analysis based on the SP\_Ace code \citep{Boeche2018}. 
The sample of MSTO-SG stars of \citet{Xiang_msto} was constructed based on LSP3  stellar parameters   (effective temperature $T_{\mathrm{eff}}$, \, surface gravity $\log\,g$, metallicity $ \mathrm{ \, [Fe/H]} $, $\alpha$-element abundance ratio  [$\alpha$/Fe], radial velocity $V_{r}$, absolute magnitudes $ M_{\rm V}$ and $ M_{\rm Ks}$ ).  Typical  uncertainties of the parameters  are listed in Table \ref{accuracy} for  SNRs higher  than  20 and  50.  Note that all the MSTO-SG sample stars of \cite{Xiang_msto} have SNRs higher than 20. Stellar distances are inferred from the distance modulus using the absolute magnitudes derived directly  from the spectra, and utilising  interstellar extinction derived with the `star pair' method \citep{yuan2013,yuan-lamost,Xiang-lss-gac}. The average distance uncertainty of the sample stars is 17 per\,cent.    Ages of the stars  are determined by  matching with
stellar isochrones using a Bayesian algorithm, as detailed in \cite{Xiang_msto}.  
The  derived ages  are shown  to be accurate to  better than  30\,per\,cent via a series of internal and external comparisons.  

In this work, we discard  MSTO-SG stars of $T_{\rm eff}\,>\,7500\,$\,K and $M_{\rm V}\,>\,4.5$\,mag as they have relatively large uncertainties in their [Fe/H] estimates, which is used to investigate the [Fe/H] distributions. Here we refer to this sample as ``MSTO-SG-SNR20''.     Moreover, the study of  [$\alpha$/Fe] distributions of mono-age stellar populations requires  higher SNRs because the [$\alpha$/Fe] measurements are more sensitive to spectral SNRs. For this purpose, we further define a sub-sample with spectral SNRs $>$\,50.  We refer this sample to ``MSTO-SG-SNR50''.  
However, the [$\alpha$/Fe] estimates from LSP3 for relatively hot (e.g.  $>$ 6000\,K) thus young stars may suffer large uncertainties due to inadequacies of the adopted template spectra as well as the intrinsic weak spectral features of $\alpha$-elements. In addition, \cite{Xiang_msto} suggest that about 10--20 per\,cent of stars younger than 4\,Gyr may be blue stragglers whose ages have been wrongly estimated. Therefore, one needs to be cautious about the derived [$\alpha$/Fe] gradients and distribution profiles of the young populations in the current work.   
Compared to the MSTO star sample used by \cite{Xiangmdf} for the study of the Galactic disc's metallicity gradients, the current MSTO-SG star sample has been significantly improved in terms of that, besides the significantly increase of sample size and the inclusion of [$\alpha$/Fe] which is not available in \cite{Xiangmdf}, systematic errors in age estimates are reduced due to improvements in stellar parameter determinations. 
As a consequence,  potential sources of contamination are also better understood and well-controlled.  
\cite{Xiangmdf}  selected MSTO in the $T_{\mathrm{eff}}$-$\log\,g$ plane (see Fig.\,1 of \cite{Xiangmdf} ), using parameters derived with the LSP3 version for the LSS-GAC DR1. While  the $\log\,g$ estimates then suffer from significant (systematic) uncertainties \citep{ren2016}. \cite{Xiang_msto} have selected the MSTO-SG stars  in the $T_{\mathrm{eff}}$-M$_{V}$ plane (see their Fig.\,5).  Values of  ${\rm M}_V$ have been derived directly from the LAMOST spectra with  a machine learning method based on kernel-based component analysis  utilising  the LAMOST-Hipparcos common stars as a training dataset,  and the systematic errors are found to be negligible \citep{Xiang_msto,Xiang-lss-gac}.   Nevertheless, we note that despite all these improvements, the current MSTO-SG sample stars may still suffer from contaminations from dwarf, giant, supergiant and binary stars due to  random errors  in the stellar parameters. The contaminations  could be  severe  and non-negligible for the old stellar populations.

\begin{table*}
\centering
\caption{Typical  accuracies of  atmospheric parameters  determined with LSP3 for stars of  spectral SNRs$>$ 20 and $>$ 50.}
 \begin{tabular}{cccccccc}

 \hline
  &  $T_{\mathrm{eff}}$ (K) & $\log\,g$ (dex) & [Fe/H] (dex) & [$\alpha$/Fe] (dex) & $V_{r}$  ($\rm km\,  s^{-1}$) & $M_{\rm V}$ (mag) & $M_{\rm Ks}$ (mag) \\
\hline
 SNRs $>$ 20 & 150 & 0.2 & 0.15 & 0.1 & 5 & 0.5 -- 0.6 & 0.5 -- 0.6 \\
 \hline
 SNRs $>$ 50 & 100 &  0.1 & 0.1 & 0.05 & 5 & 0.2 -- 0.3 & 0.2 -- 0.3 \\
  \hline
 \end{tabular}
 \label{accuracy}
 \end{table*}
   
\subsection{Selection effect correction}
 \cite{Nandakumar2017}  and \cite{Chen2018} have suggested that the LAMOST selection function has only a marginal effect on the
the peak of the metallicity distribution. Nevertheless, both \cite{Nandakumar2017} (see their Fig.\,10)  and \cite{Chen2018} (see their Fig.\,11) do find that the dispersion and skewness of the MDF are affected  by the selection effects of  LAMOST. The possible effects of the selection effects of LAMOST on the MDF of mono-age stellar populations  remain to be investigated.  To clarify the situation, we have considered and corrected for the selection function of our sample stars. 

Generally, the selection effects of our sample concerned in the current work mainly consist of  two parts: (1) the selection effects of the LAMOST Galactic spectroscopic surveys with respect  to the photometric catalogues from which the spectroscopic targets are  selected. The latter is usually assumed to be complete (to a certain limiting magnitude) and represents  the underlying stellar populations; and (2) the incompleteness of MSTO-SG stars in a given volume   caused by  the limiting magnitude of the  surveys and the wide range of absolute magnitudes of MSTO-SG stars.  Such an effect also varies with stellar populations of different ages, as they have different intrinsic width of absolute magnitude range according to their definition \citep{Xiang_msto}.  The first part is corrected using the selection function $S$
 derived by   \cite{Chen2018}. Based on the source of origin, $S$ contains  also two portions. The first portion, quantified by $S_{1}$, characterises the LAMOST target selection strategy. The second portion, quantified by $S_{2}$,  characterises the selection effects due to the observational quality, data reduction and parameter determination.   Each star in our sample is weighted  by  $\frac{1}{S}$ in the colour-magnitude diagram (CMD). 
 In order to account for  the volume incompleteness of the mono-age populations of MSTO-SG stars,  we define the   completeness  volumes of  our sample stars  using  the same method of \cite{Xiang2018} and \cite{ChenBQ}.  The complete  volume of a sample of stars is a combined  consequence of the  absolute magnitude, the limiting magnitude of the LAMOST Galactic spectroscopic surveys and the dust extinction. Considering the absolute magnitudes of the MSTO-SG stars vary significantly with age,  we define a complete volume for each of the  mono-age  stellar populations of  age range $0.0 < \tau < 2.0$,  $2.0 < \tau < 4.0 $, $4.0 < \tau < 6.0 $,  $6.0 < \tau < 8.0 $, $8.0 < \tau < 11.0 $,  $11.0 < \tau < 14.0$  and $0 < \tau < 14.0$ \,Gyr, respectively.   Sample stars falling  outside  the defined complete volume are discarded. 
The final completed samples of MSTO-SG-SNR20 and MSTO-SG-SNR50 contain 453,188 and 291,126 stars, respectively. 

\begin{figure}
\centering
\includegraphics[width=3.5in]{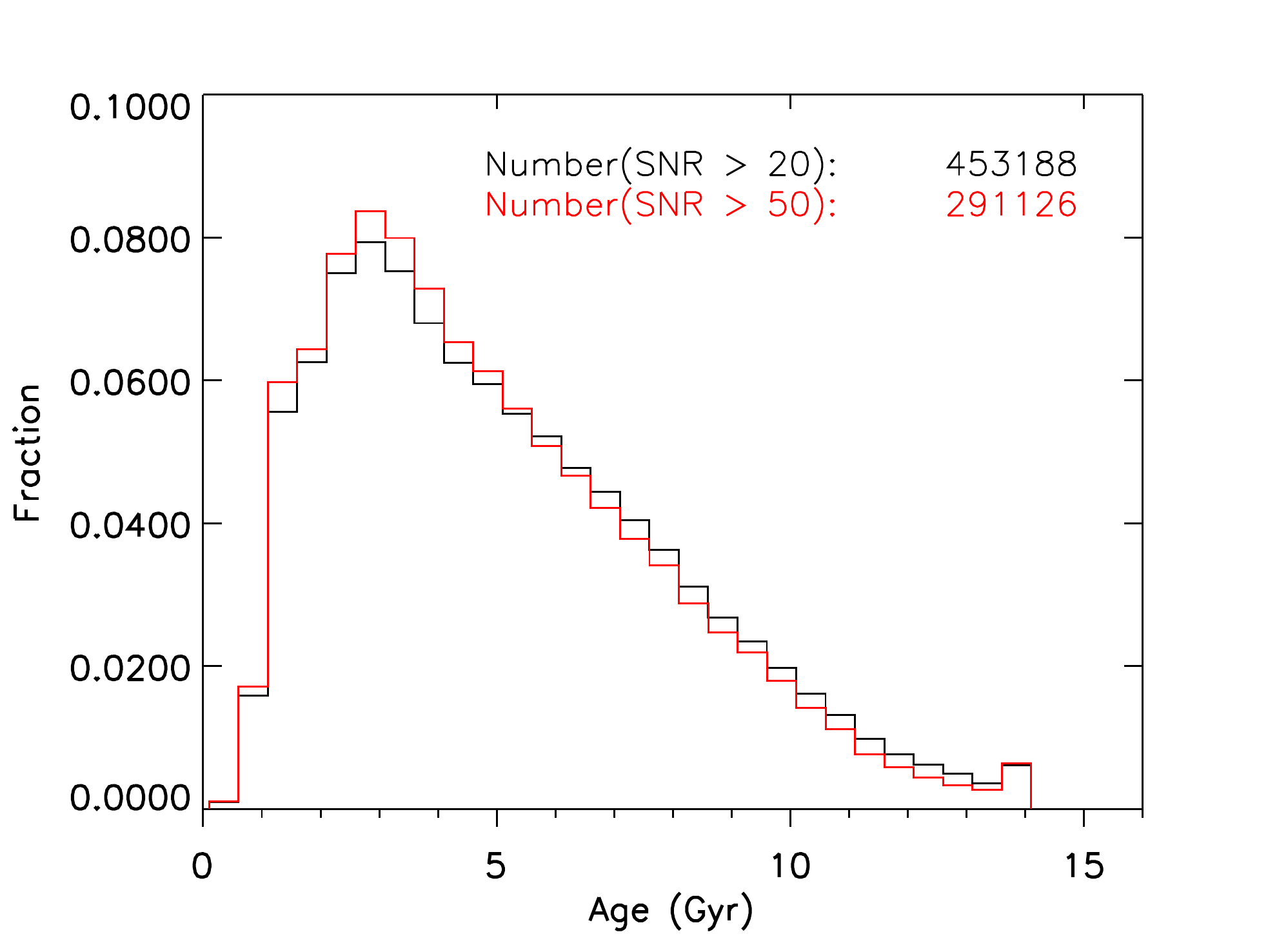}
\caption{Normalised  age distributions  of the MSTO-SG-SNR20  (black line) and MSTO-SG-SNR50 (red line)  samples. The total numbers of stars in the two samples are labeled in the Figure.}
\label{age_distribution}
\end{figure}
 
\begin{figure*}
\centering
\includegraphics[width=7.0in]{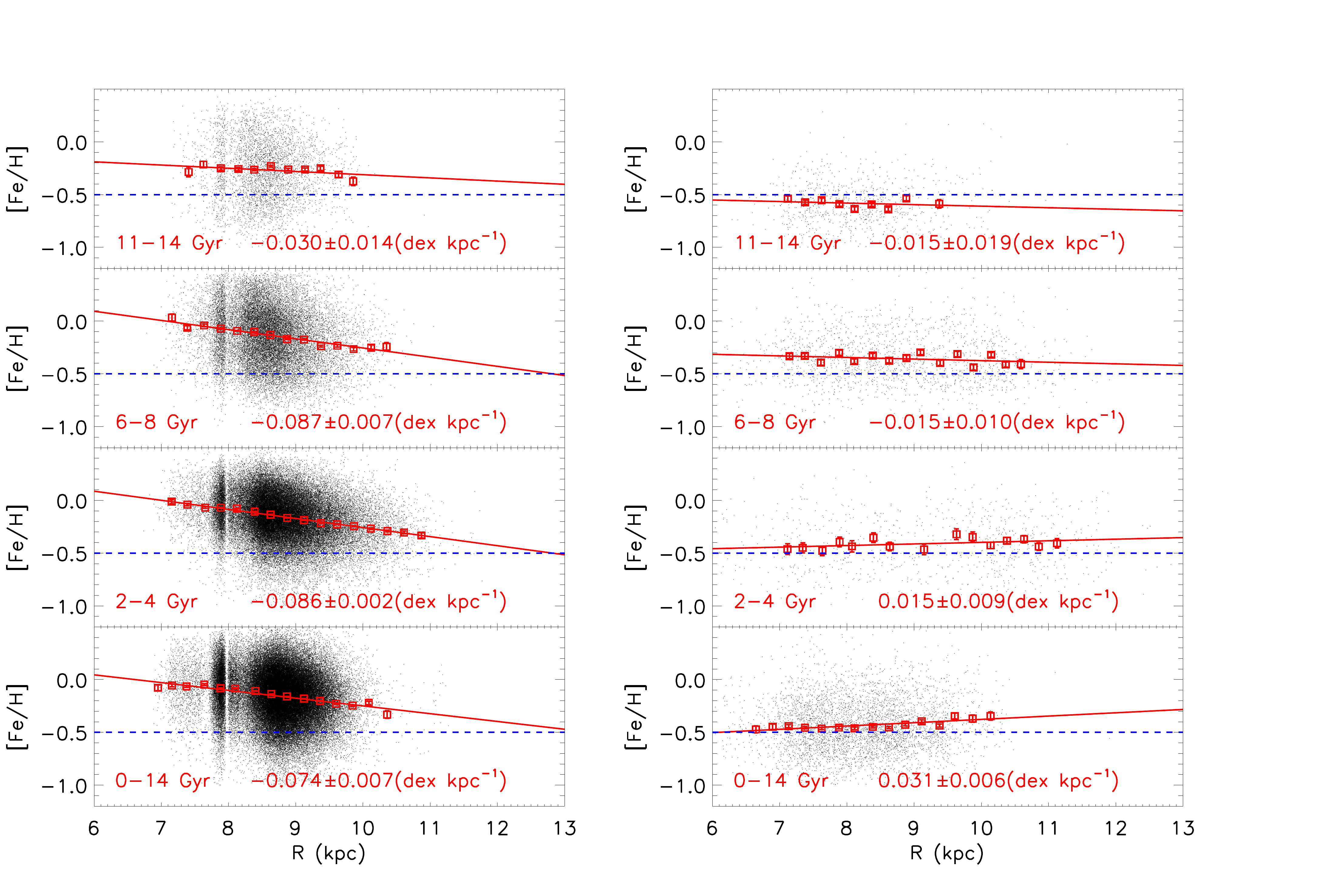}

\caption{Radial [Fe/H] gradients derived from several mono-age stellar populations (as marked in each panel) in  two height slices: $|Z| < 0.3$ kpc (left panels),  $1.5 < |Z| < 2.0$ kpc (right panels).  Red squares represent the CMD-weighted mean values of [Fe/H] of the individual radial bins.  The error bars of the mean values are also plotted, but in almost all the cases they are smaller than the size of the symbols. Linear fits to the red squares are over plotted by red lines.   The blue  horizontal  dashed lines  delineate a constant [Fe/H] value of $-0.5$\,dex.  The age span of each population,  the slope of the linear fit (the radial [Fe/H] gradient) and the associated error,  are marked in each panel. The mean value of [Fe/H] as a function of $R$ in each panel can be well fitted by a linear line.}

\label{feh_gradients}
\end{figure*}

\begin{figure*}
\centering
\includegraphics[width=7.0in]{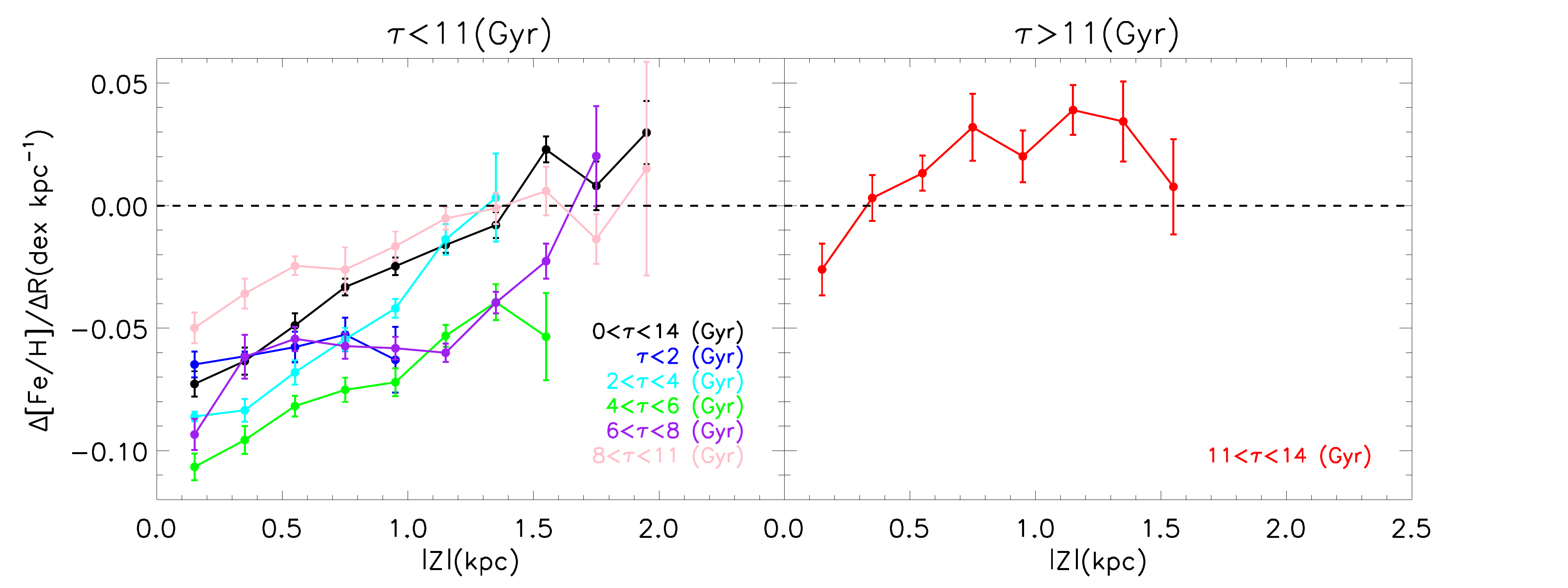}

\caption{Radial [Fe/H] gradients as a function of $|Z|$ derived from mono-age  populations, presented with different colours, as marked in the plot. The horizontal dashed lines delineate a constant radial [Fe/H] gradient value of 0.0\,dex\,$\rm kpc^{-1}$. The mean radial [Fe/H] gradients for the young (left panel) and the old (right panel) stellar populations are negative and positive, respectively. }

\label{feh_gradients_r}
\end{figure*}

\begin{figure}
\centering
\includegraphics[width=3.5in]{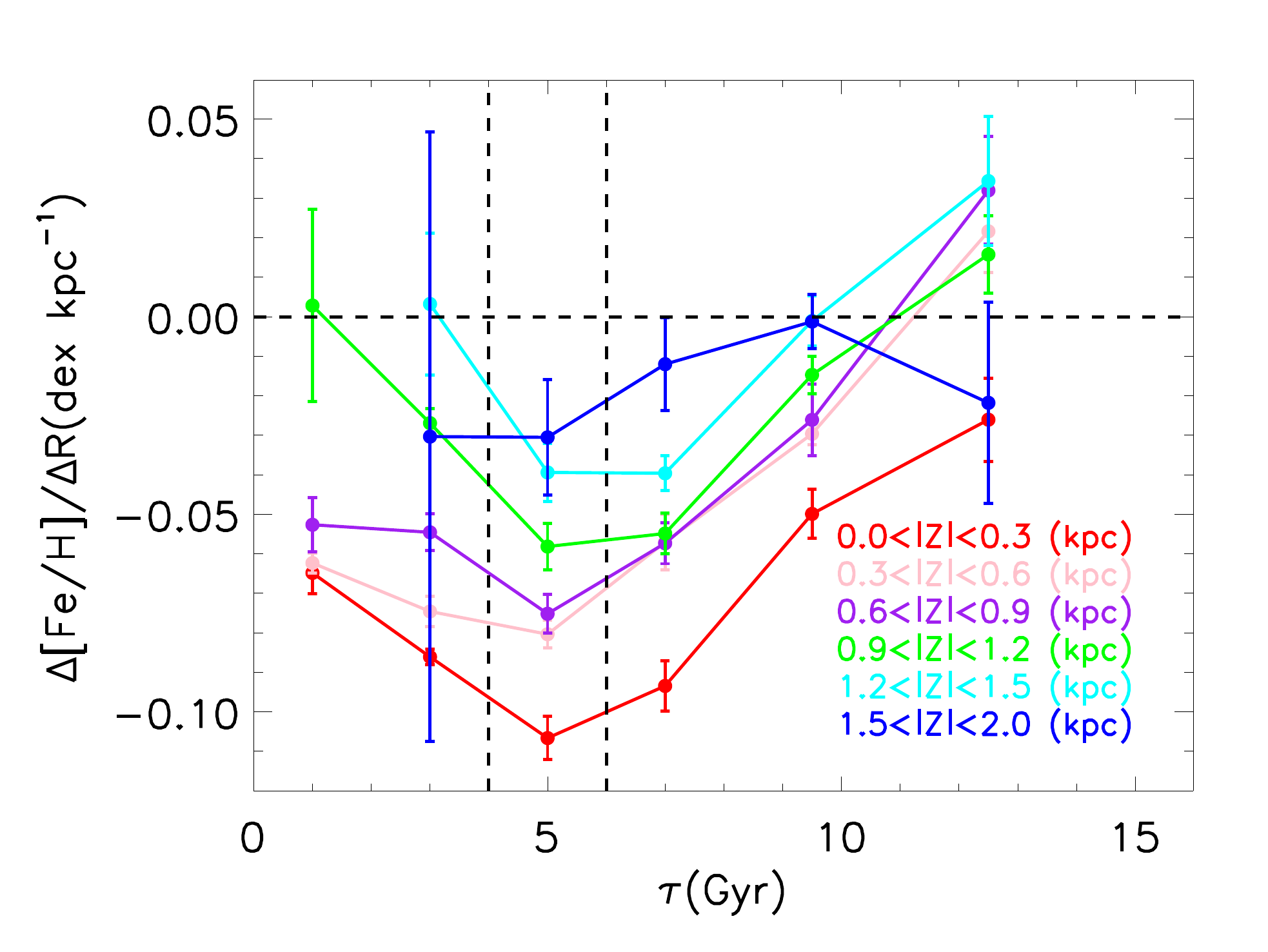}

\caption{Radial [Fe/H] gradients as a function of $\tau$ derived from stars at different $|Z|$ slices, presented with different colours, as marked in the plot. The horizontal dashed line delineates a constant radial [Fe/H] gradient  of 0.0\,dex\,$\rm kpc^{-1}$. The vertical  dashed lines delineate constant ages   of 4  and 6 Gyr. There are dips in the radial [Fe/H] gradients at 4$\,<\,\tau\,<\,$6\,Gyr. }

\label{feh_gradients_age}
\end{figure}

\begin{figure*}
\centering
\includegraphics[width=7.0in]{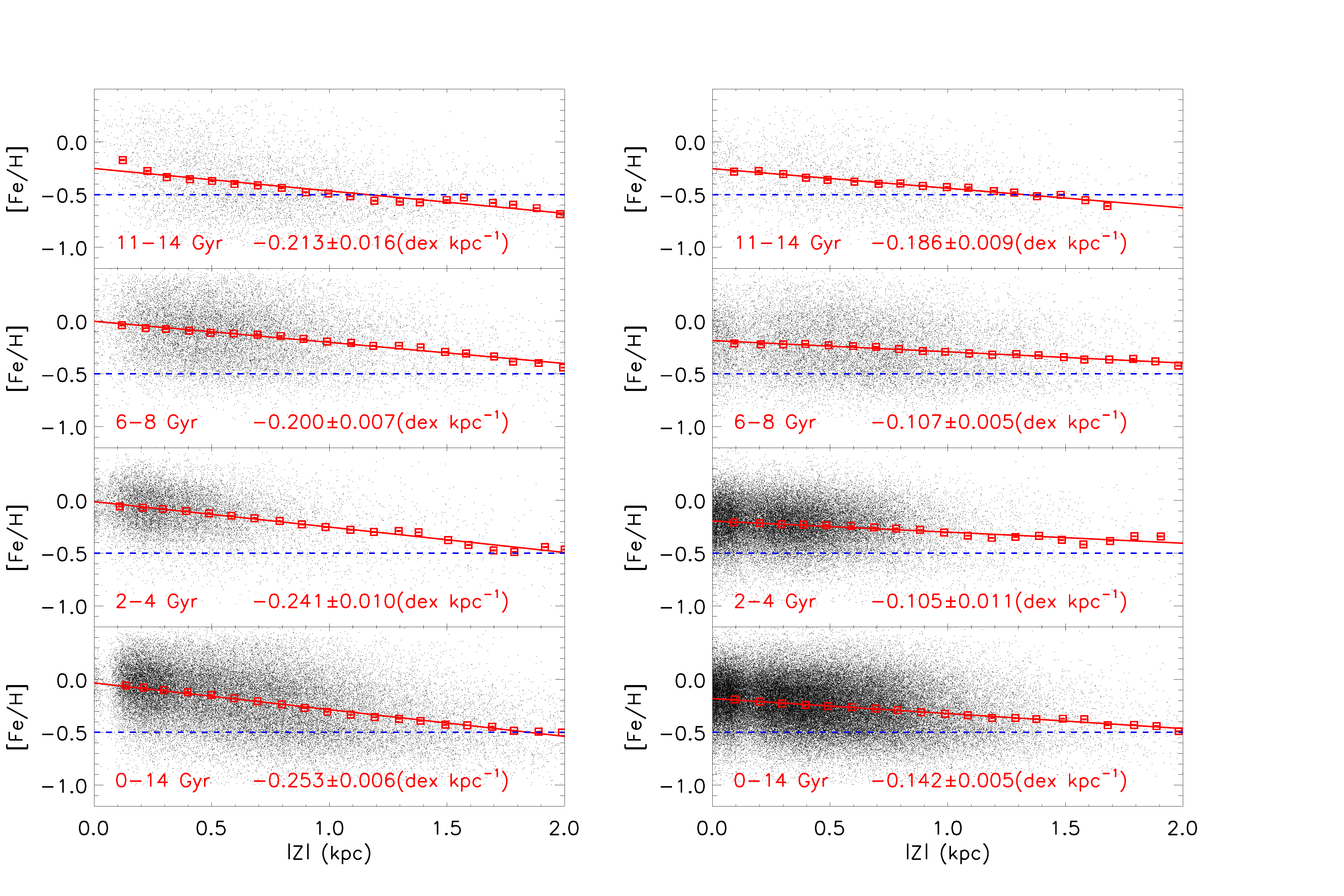}
\caption{Vertical  [Fe/H] gradients derived for several individual  mono-age  stellar populations (as marked in each panel)  in  two radial bins: $7 < R < 8$ kpc (left panels) and $9 < R < 10$ kpc (right panels).  Red squares represent the CMD-weighted mean values of [Fe/H] in the individual vertical bins.  The error bars of the mean values are also over plotted, but in almost all the cases they are smaller than the size of the symbols. Linear fits to red squares are over plotted by red lines. The blue  horizontal  dashed lines  delineate a constant [Fe/H] value of $-0.5$\,dex. The age range of each mono-age  stellar population,  the slope of the linear fit (vertical [Fe/H] gradients) and the associated error are marked in each panel. The mean value of [Fe/H] as a function of $|Z|$ in each panel  can be well fitted by a straight line.}

\label{feh_gradient_r}
\end{figure*}

Fig.\ref{spatial_distribution} shows the distributions of  [Fe/H], [$\alpha$/Fe] and  number density  of the final  (complete) MSTO-SG-SNR20  and MSTO-SG-SNR50  samples in the $R$--$Z$  plane.  
Here $R$ and $Z$ are the axes of the Galactocentric cylindrical coordinate system ($R$, $\Phi$, $Z$).  The Sun is assumed to be at the Galactic disc mid-plane (i.e. $Z=$\,0\,pc) and has  a $R$ value  of 8\,kpc. 
Fig.\ref{age_distribution} shows the normalised  age  distributions of the final MSTO-SG-SNR20   and  MSTO-SG-SNR50   samples.  The distributions show that for both samples, the stars have a broad distribution in stellar age, from $<1$\,Gyr to $>13$\,Gyr, and the distributions peak at 3--4\,Gyr.
 
\begin{figure}
\centering
\includegraphics[width=3.5in]{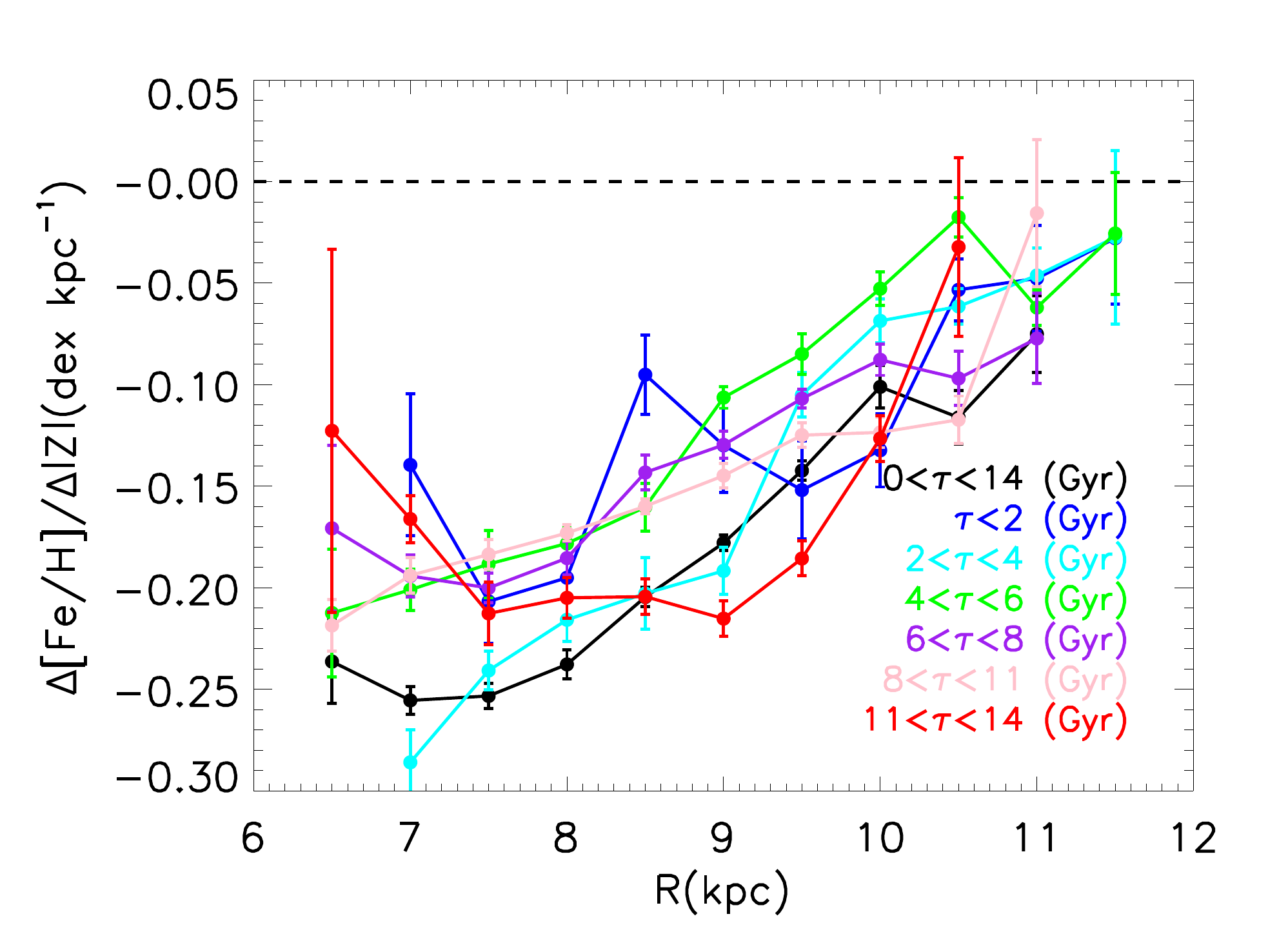}

\caption{Vertical [Fe/H] gradients as a function of $R$ derived from mono-age stellar populations, presented with different colours, as marked in the plot. The horizontal dashed line delineates a constant vertical [Fe/H] gradient of $0.0$\,dex\,$\rm kpc^{-1}$. Stars in all $R$ and $\tau$ bins have negative vertical [Fe/H] gradients.}

\label{feh_gradients_z}
\end{figure}

\begin{figure*}
\centering
\includegraphics[width=7.0in]{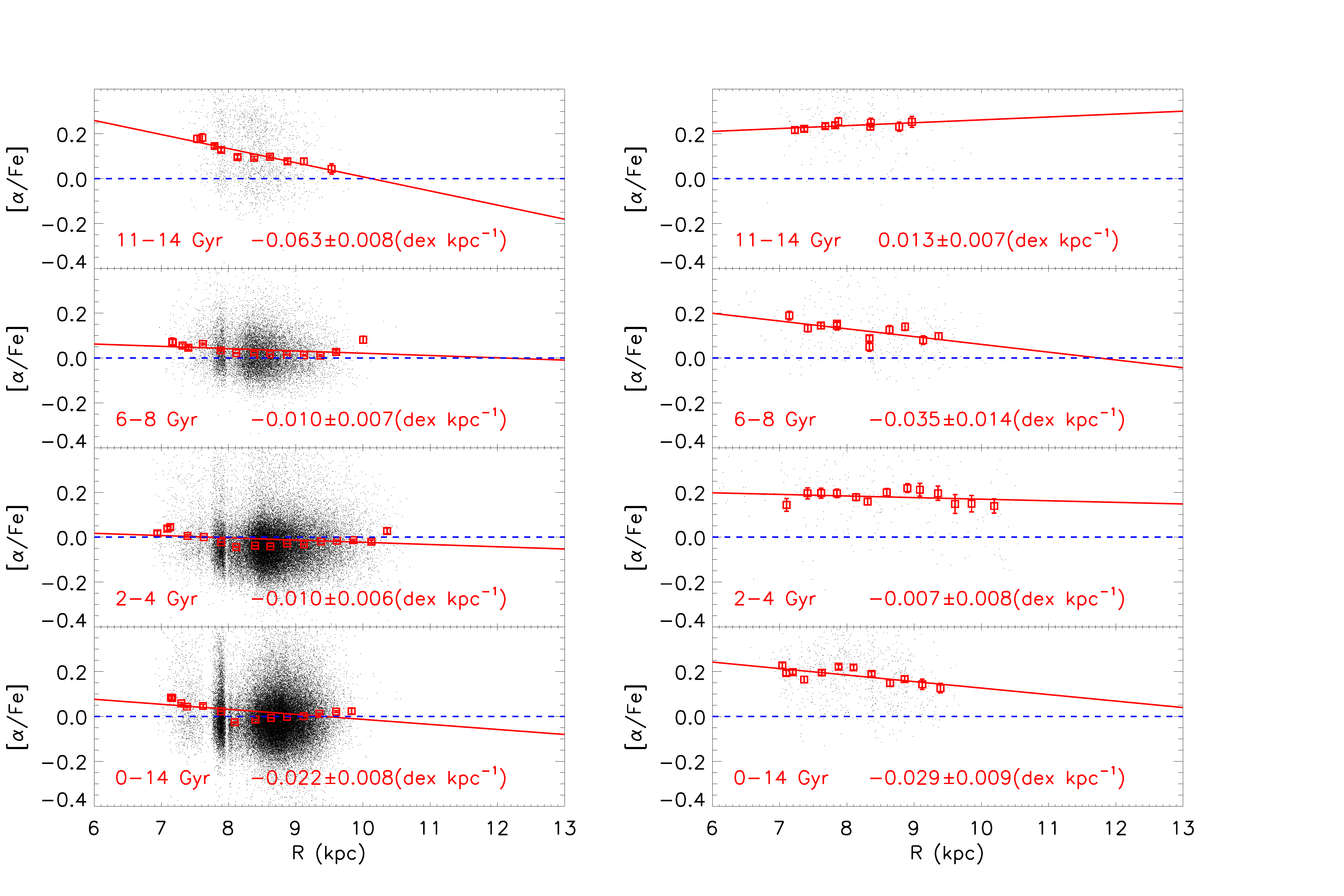}

\caption{Radial [$\alpha$/Fe] gradients derived from several mono-age stellar populations (as marked in each panel) in  two height slices: $|Z| < 0.3$ kpc (left panels) and  $1.5 < |Z| < 2.0$ kpc (right panels).  Red squares represent the CMD-weighted mean values of  [$\alpha$/Fe] of the individual radial bins.  The error bars of the mean values are also plotted, but in almost all the cases they are smaller than the size of the symbols. Linear fits to the red squares are over plotted by red lines. The blue  horizontal  dashed lines  delineate a constant  [$\alpha$/Fe] value of $0.0$\,dex. The age span of each population,  the slope of the linear fit (the radial [$\alpha$/Fe] gradient) and the associated error,  are marked in each panel. In some $|Z|$ and $\tau$ bins, [$\alpha$/Fe] as a function of $R$ can not be well fitted by a linear line.}

\label{afe_gradients}
\end{figure*}

 \section{Radial and vertical gradients  of [Fe/H] and [$\alpha$/Fe] }
 In this Section, we investigate the radial and vertical gradients of [Fe/H] with the MSTO-SG-SNR20 sample and the radial and vertical gradients of  [$\alpha$/Fe]  with  the MSTO-SG-SNR50 sample using a method similar to \cite{Xiangmdf} in order to better compare with their results. 
 \subsection{The radial [Fe/H]   gradients}
 Firstly, we determine the radial [Fe/H] gradients of mono-age stellar populations in several thin slices of   $|Z|$.  
 In doing so,  we divide the stars in the   MSTO-SG-SNR20 sample  into mono-age  populations of 6 age bins: $\tau <$ 2.0, $2.0 < \tau < 4.0$, $4.0 <\tau < 6.0$, $6.0 < \tau < 8.0$, $8.0 < \tau < 11.0$, $11.0 < \tau < 14.0$ Gyr.  The division of age bins is the same as \cite{Xiangmdf}, which allows a direct comparison with their results. Further we divide each of  the mono-age  populations into bins of a  constant thickness of  $\Delta |Z| =$ 0.3 kpc and 0.5\,kpc for stars of  $|Z| <$\,1.5 kpc and   $|Z| >$\,1.5 kpc, respectively.
 In each age and height bin, we further divide the stars into small  radial annuli of 0.25 kpc. 
Bins containing less than 30 stars are discarded. 
A straight line is then adopted to fit the CMD-weighted mean values of [Fe/H] in the  individual radial annuli, as a function of $R$. The slope of the line is adopted as the radial [Fe/H] gradient.  Fig.\ref{feh_gradients} plots the results for  two $|Z|$ slices of several individual  mono-age  stellar populations: $|Z| < 0.3$ kpc and $1.5 < |Z| < 2.0$ kpc.    It shows that the CMD-weighted mean values of [Fe/H] as a function of $R$ can be well fitted  by a straight line.   The dip in the number of stars at $R\,\sim\,8$\,kpc and $|Z|\,<\,0.3$\,kpc in the Figure is a consequence of the bright limiting magnitude ($\sim$\,9\,mag in $r$-band) of the survey.

 Fig.\ref{feh_gradients_r} shows the derived radial [Fe/H] gradients as a function of $|Z|$ for mono-age stellar populations.  It shows that there are negative radial [Fe/H] gradients for stellar populations of $\tau < 11$ Gyr at  $ |Z| <$ 1.5 kpc.   The negative radial metallicity gradients flatten with increasing $|Z|$.  
 The gradients of stellar populations of $\tau >$ 11 Gyr have marginally  positive values except the results of  $|Z| < 0.3$\,kpc ($\sim$ $-$0.03$\pm$0.04 dex $\rm kpc^{-1}$), and vary little with  $|Z|$.  
 
 The   values of radial [Fe/H] gradients  presented in the current paper are slightly  larger than those of \cite{Xiangmdf} by  0.0 -- 0.05\,dex\,$\rm kpc^{-1}$. The differences are larger at  low $|Z|$ bins.  As discussed in Section\,2.1, compared to  \cite{Xiangmdf},  ages, distances and  atmospheric parameters of stars in the current sample have been improved. The current work also adopts different criteria  of selecting  MSTO-SG (or simply MSTO) compared to \cite{Xiangmdf}. 
We re-estimate the radial [Fe/H] gradients using the common stars of  the current sample and that  used by \cite{Xiangmdf} to check whether the difference of the parameters are mainly responsible for the difference.  The exercise yields very similar results, suggesting  that  the differences in the derived parameters are not mainly responsible for the differences of  the results of  radial [Fe/H] gradients derived in the current work and those presented in \cite{Xiangmdf}. A further inspection of the MSTO sample of \cite{Xiangmdf}  using the updated set of stellar parameters suggests that a considerable fraction of the sample stars are actually main-sequence star contaminations due to systematic errors in the (old version) LSP3 log\,$g$ estimates, which are used by \cite{Xiangmdf} to select the MSTO sample stars. The contamination rate is a function of $T_{\rm eff}$ in the sense that the cooler  the stars, the higher the fraction of the contaminations.  Because of  the  magnitude limits  of the LAMOST surveys, the closer to the Sun, the cooler and  more metal-rich stars  could be observed by LAMOST  as they are intrinsically fainter. Besides, most of the Galactic radii of the MSTO stars in \cite{Xiangmdf} are  larger than 7.5\,kpc (see their Fig.\,13), the radial gradients  are derived with the stars of $R > 7.5$\,kpc (see their Fig.\,14). Thus, we can approximately think that there are more  contaminations (by more metal-rich stars)  in the  inner disc than in the outer disc, which produce the steeper radial [Fe/H] gradients in  \cite{Xiangmdf}. Also, the contaminations of metal-rich stars are more severe at lower heights, which causes more overestimation (steeper) of radial [Fe/H] gradients at lower disc.   So that in conclusion, the differences of radial [Fe/H] gradients between this work and \cite{Xiangmdf} are likely caused by overestimation of the results derived by \cite{Xiangmdf}  due to non-negligible main-sequence star contaminations in their MSTO sample. 
 
The  vertical variations of radial [Fe/H] gradients are quite  similar to what found by  \cite{Xiangmdf}  as shown in their Fig.\,15.   \cite{Minchev2014} find that the negative radial metallicity gradients  strongly flatten with increasing $|Z|$ in their chemodynamical models (see their Fig.\,10). This  is consistent with our results for stellar populations of $\tau < 11$\, Gyr.

Fig.\ref{feh_gradients_age} shows the derived radial [Fe/H] gradients as a function of age for stars at different $|Z|$ bins. It shows that  for  stellar populations of $\tau < 4$ Gyr, the gradients steepen with increasing age, reaching a maximum at $4 < \tau < 6$ Gyr, and then  flatten with age.

The existence of reversions  of radial [Fe/H] gradients as a function of age are also found by \cite{Xiangmdf} (see their Fig.\,16). However, they found the reversion occurs at $6 < \tau < 8$\,Gyr. As we  know the stellar age used by \cite{Xiangmdf} are systematically  older than those  deduced by \cite{Xiang_msto} for stars of  $\tau>$\,6\,Gyr.  This accounts for the difference  of the reversion epoch. 
Interestingly, the epoch of reversion derived here is closer to   that obtained by \cite{Casagrande} ($\sim$ 4 Gyr) for the solar-neighbourhood disc ($Z < 100$\,pc) using GCS FG(K) dwarf stars, as shown in their Fig.\,18.  The results are not consistent with that obtained by the models of \cite{Toyouchi2018}, who find that the radial [Fe/H] gradients monotonically increasing as age increases. The disagreement may be the response of their too simple and perhaps unrealistic assumptions for gas infall, gas outflow, gas reaccretion and the stellar radial migration processes, or lack of some important ingredients for an understanding of the MW formation \citep{Toyouchi2018}.

  \subsection{The vertical [Fe/H]   gradients}
To derive the vertical [Fe/H] gradients, we divide the stars in the   MSTO-SG-SNR20 sample into radial annuli of 1 kpc with a step of 0.5\,kpc in the radial direction for the mono-age stellar  populations. In each annulus, we further divide the stars into bins of  constant height  of $\Delta |Z|=$ 0.2 kpc with a step of 0.1 kpc. Bins containing less than 30 stars are discarded. The  CMD-weighted mean values of [Fe/H] in the individual  $|Z|$ slices as a function of  $|Z|$ are fitted by straight  lines to derive  the vertical [Fe/H] gradients.  Fig.\,\ref{feh_gradient_r} plots the results for two radial bins of several mono-age stellar populations: $7 < R < 8$ kpc and $9 < R < 10$\,kpc.  It shows that  CMD-weighted mean values of [Fe/H] as a function of $|Z|$ can  be well fitted   by a straight  line.  Fig.\,\ref{feh_gradient_r}  shows that  old stars reach much further above the disc than young stars.  Few stars are sampled at $7\,<\,R\,<\,8$\,kpc and $|Z|\,\sim\,$0\,kpc because of the bright limiting magnitude of the surveys. 


Fig.\ref{feh_gradients_z}  shows the derived vertical [Fe/H] gradients as a function of $R$ for the  mono-age stellar populations.  It shows that almost all the   mono-age stellar populations have negative vertical [Fe/H] gradients. The vertical [Fe/H] gradients flatten with increasing $R$ for all the mono-age stellar populations at  $R > 8$\,kpc.  The vertical [Fe/H] gradients vary significantly with age.   

Compared to \cite{Xiangmdf}, we have extended  the measurement of vertical [Fe/H] gradients toward the inner disc to $R\,<\,7$\,kpc  and found that the vertical [Fe/H] gradients in the inner disc are much flatter than that in the solar neighbourhood.  The   vertical [Fe/H] gradients derived  in the current work are  consistent with those obtained by  \cite{Xiangmdf} for populations  younger than 11 Gyr.  The derived vertical [Fe/H] gradients for the  oldest  populations ($\tau >$ 11 Gyr) are  comparable to those for  $8 < \tau <11$ Gyr in the current work. Whereas \cite{Xiangmdf} found much shallower gradients for their oldest populations than those for $8 < \tau <11$ Gyr.  We believe that this is because \cite{Xiangmdf}  discard stars cooler than 5,400\,K.  Such a temperature cut may have   rejected  some old metal-rich stars which are mostly found  at low $|Z|$'s, leading to  shallower vertical [Fe/H] gradients for the populations of $\tau >$\,11\,Gyr.   Of course, it is also possible that those old metal-rich MSTO-SG stars in the current  sample may be contaminations from  younger dwarf stars.  A further study is needed to deeper understand the results.   The cause of the differences in  the vertical [Fe/H] gradients of the oldest stellar populations ($\tau >$ 11 Gyr)  needs to be  further  explored.

In the current work,  we do not find  a monotonously increase of vertical [Fe/H] gradient with age, in contrast to   \citet[][Fig.\,17]{Xiangmdf}, who find that (with the exception of the oldest stars of $\tau\,>\,$11\,Gyr)  the younger the stars, the flatter the vertical [Fe/H] gradients.  \citet[][Fig.\,6]{Ciuca2018}  also find a  clear trend of vertical [Fe/H] gradient with age using common dwarf stars selected from  Gaia and RAVE, quite similar to what \cite{Xiangmdf} find.  
We note that the observational selection bias of the samples of \cite{Ciuca2018} and \cite{Xiangmdf} are not removed completely.  \cite{Ciuca2018} only minimise the observational selection bias induced by the
various selection criteria by selecting a sample in a specific colour
and magnitude range. \cite{Xiangmdf} did not consider the effects of incomplete volume of mono-age stellar populations on the derived vertical [Fe/H] gradients. The selection bias may be responsible for the difference of the trend between our results and that of \cite{Ciuca2018} and \cite{Xiangmdf}. 

 \subsection{The radial  [$\alpha$/Fe] gradients}
We  determine  the radial [$\alpha$/Fe] gradients in  thin $|Z|$ slices. In doing so,  the stars in the MSTO-SG-SNR50 sample are divided into different $|Z|$ and age bins as described in Section\,3.1.   In each bin, stars are further divided into small radial  bins of size of 0.25 kpc. Bins containing less than 30 stars are again discarded. The CMD-weighted mean values of $[\alpha$/Fe] in the individual  radial bins  as a function of $R$ are fitted by a straight line to derive the radial [$\alpha$/Fe] gradient.  As an example, Fig.\,\ref{afe_gradients} presents the  results for several mono-age populations at the  two $|Z|$ slices: $|Z| <$\,0.3\,kpc and\,$1.5 < |Z| < 2.0$ kpc .    It  shows that   straight lines fit the trends  well for almost all the mono-age stellar populations of  almost all the $|Z|$ bins.  An exception is for  young stellar populations ($\tau < 4$ Gyr), and this  may be a consequence of  poor  [$\alpha$/Fe] estimates for those stars \citep{Xiang_msto}.  For the  population of all ages ($0 < \tau < 14$ Gyr), the fits are also not very good, and this may be the consequence of  the dominations of young disc stars with incorrect [$\alpha$/Fe] values  at low $|Z|$'s of the inner disc. 

\begin{figure*}
\centering
\includegraphics[width=7.0in]{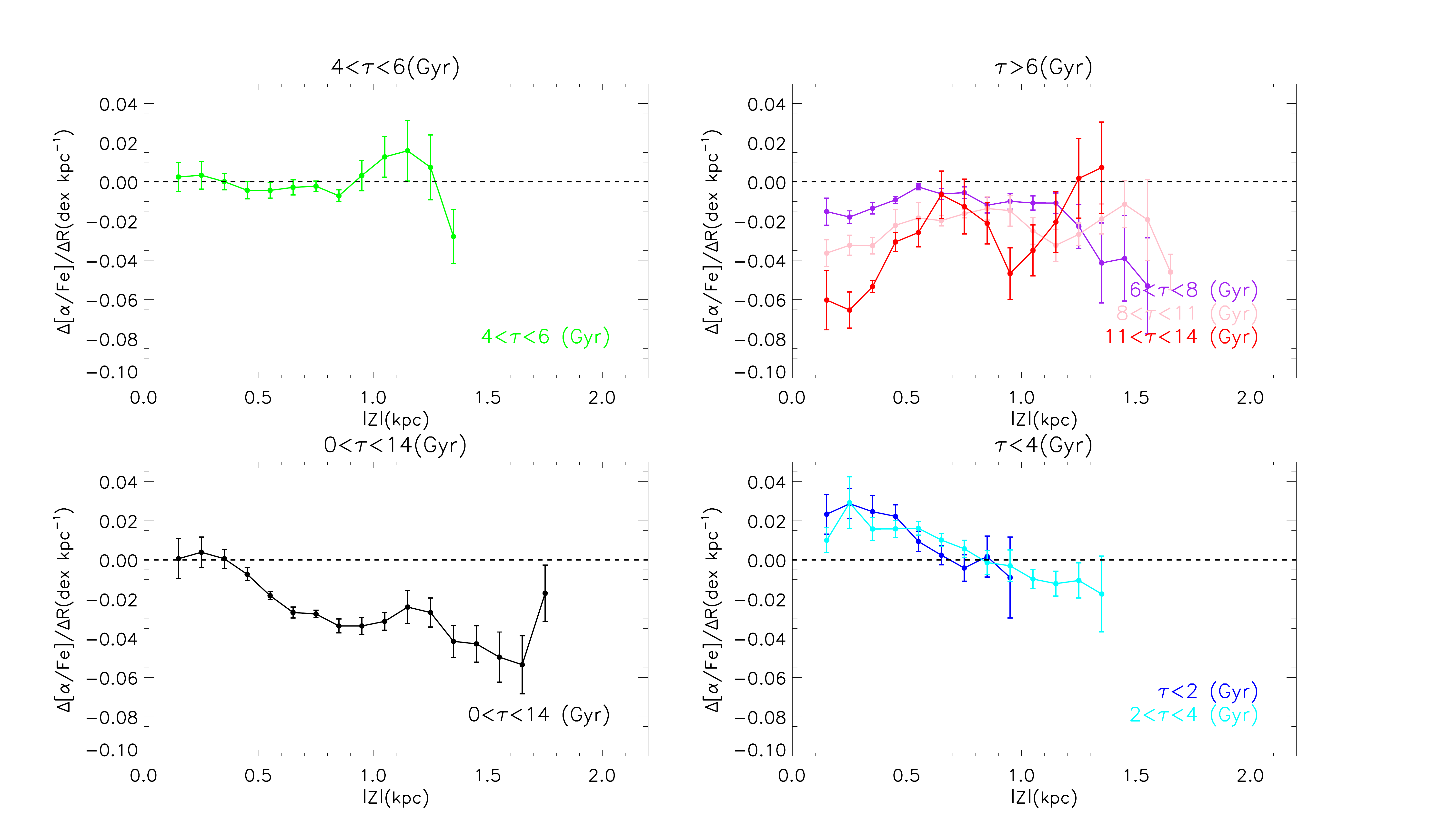}
\caption{Radial [$\alpha$/Fe] gradients as a function of $|Z|$ derived from mono-age populations, presented with different colours, as marked in the plot. The mean radial  [$\alpha$/Fe] gradients of all ages, young, median age and old stellar populations are negative, positive, near-zero and negative, respectively.  The horizontal dashed  lines delineate a constant radial [$\alpha$/Fe]  gradient value of 0.0 dex $\rm kpc^{-1}$. The results of young stellar populations should be taken with caution as about 10--20\,per\,cent  [$\alpha$/Fe] measurements of those stars  might be incorrect. }  

\label{afe_gradient_z}
\end{figure*}

Fig.\ref{afe_gradient_z} shows the derived  radial [$\alpha$/Fe] gradients for the  mono-age stellar populations as a function of $|Z|$. The current work investigate the  radial [$\alpha$/Fe] gradients of mono-age stellar populations for the first time and find  that the radial [$\alpha$/Fe] gradients and its variations with $|Z|$ vary significantly with age.   Fig.\ref{afe_gradient_z}  shows that the radial [$\alpha$/Fe] gradients of  population of all ages  are nearly flat   at low $|Z|$'s, and negative at high $|Z|$'s.  The radial [$\alpha$/Fe] gradients of  populations of $\tau < 4$ Gyr  have  positive values  at low $|Z|$'s, and negative at high $|Z|$'s, and  become smaller as $|Z|$ increases. Stars  of ages $4 < \tau < 6$ Gyr have nearly flat  radial [$\alpha$/Fe] gradients.   Stars older than  $\tau > 6$ Gyr have negative radial [$\alpha$/Fe] gradients.  The radial [$\alpha$/Fe] gradients  of old stellar populations  ($\tau > 8$ Gyr) flatten as $|Z|$ increases.

 Several previous works have studied the radial [$\alpha$/Fe] gradients of the population of all ages.  \cite{alpha}  show that there are   almost flat  radial [$\alpha$/Fe] gradients at $Z_{\rm max} < 0.4$ kpc, but  clear negative  [$\alpha$/Fe] gradients at $1.5 < Z_{\rm max} < 3.0$ kpc, derived from an analysis of  APOGEE red giant stars as shown in their Fig.\,16.     \cite{Minchev2014}  have also found  that the weak positive [Mg/Fe] gradient for the all-age stellar population turns negative as $|Z|$ increases (see their Fig.\,10). \cite{Boeche2014} have found that the radial [Mg/Fe], [Al/Fe] and [Si/Fe] gradients are positive at low $|Z|$'s,  and negative at high $|Z|$'s using giant stars selected from RAVE (see their Table\,2). Our current result of the stellar population of all ages are consistent with those of \cite{alpha}  and  \cite{Minchev2014}, but are slightly different compared to those by \cite{Boeche2014}. 

\begin{figure*}
\centering
\includegraphics[width=7.0in]{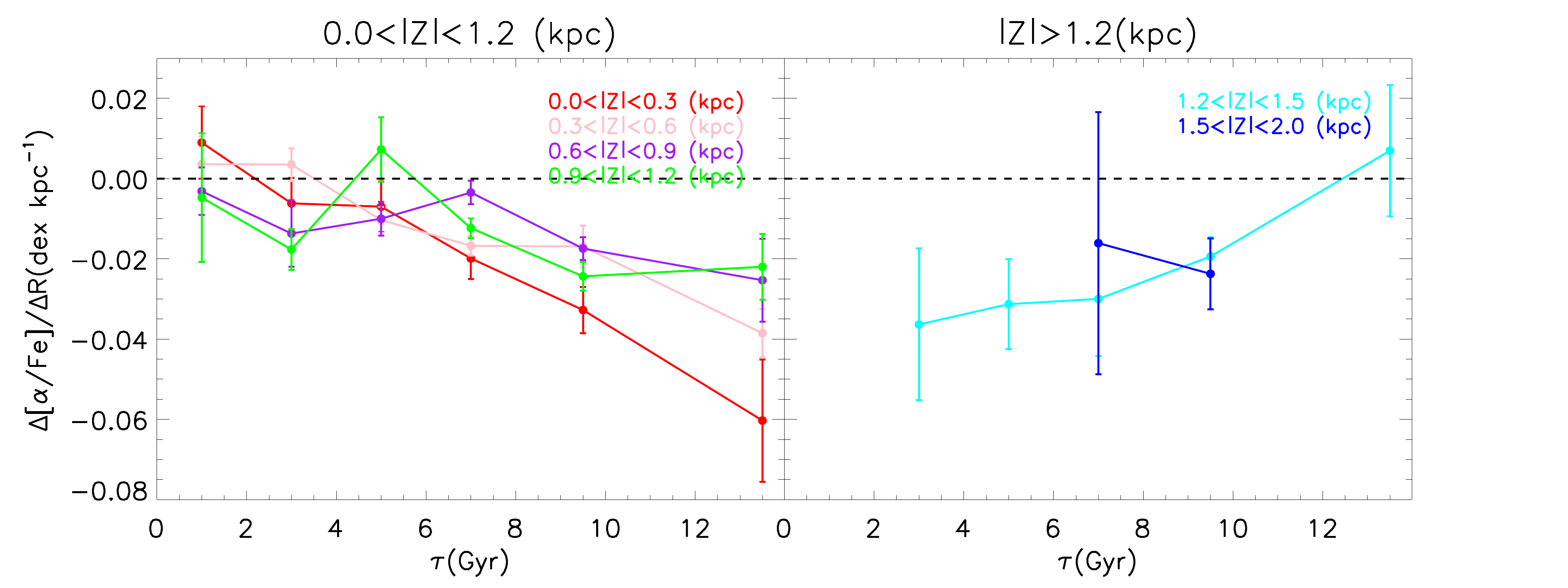}

\caption{Radial [$\alpha$/Fe] gradients as a function of $\tau$ derived for stars at different $|Z|$ slices, presented with different colours, as marked in the plot. The horizontal dashed lines delineate a constant radial [$\alpha$/Fe] gradient value of 0.0\,dex\,$\rm kpc^{-1}$.  Radial [$\alpha$/Fe] gradients steepen with increasing age at low $|Z|$ bins (left panel), but flatten or become nearly invariant with increasing age at high $|Z|$ bins (right panel). }

\label{afe_gradients_age}
\end{figure*}

Fig.\ref{afe_gradients_age} shows the derived  radial [$\alpha$/Fe] gradients as a function of age for stars at different $|Z|$'s.  It shows that the derived negative radial [$\alpha$/Fe] gradients steepen with  increasing age  at low $|Z|$'s, but flatten or become  nearly  invariant  with increasing age  at high $|Z|$'s. \cite{Minchev2014}  have found that there is  a nearly flat  [Mg/Fe] gradient for old stellar populations, but a positive  radial [Mg/Fe] gradient for young stellar population (see their Fig.\,10) in their chemodynamical models,  some different from what found here.  Our results may strongly constrain the existing chemodynamical models.

\subsection{The vertical [$\alpha$/Fe]   gradients}

We now determine  the vertical [$\alpha$/Fe] gradients.  In doing so, we divide the stars  in the MSTO-SG-SNR50 sample into  annuli of 1.0 kpc with a step of 0.5 kpc in the radial direction for the mono-age stellar populations.   In each annulus, we further divide the stars into bins of a  constant thickness of $\Delta |Z| = 0.2$ kpc with a step of 0.1 kpc. Bins containing less than 30 stars are discarded.   For each annulus, the CMD-weighted mean values of  [$\alpha$/Fe] are estimated for  all height slices. The results  as a function of $|Z|$ are fitted by a straight line. The slope of   the straight line  is adopted as the vertical gradient of the  mono-age stellar population at  that radial annulus.  Fig.\ref{afe_gradients_r} shows the results   at two radial bins: $7 < R < 8$ kpc and $9 < R < 10$\,kpc.    It shows that the CMD-weighted mean values of [$\alpha$/Fe] as a function of $|Z|$ are well fitted  by a linear line.   

\begin{figure*}
\centering
\includegraphics[width=7.0in]{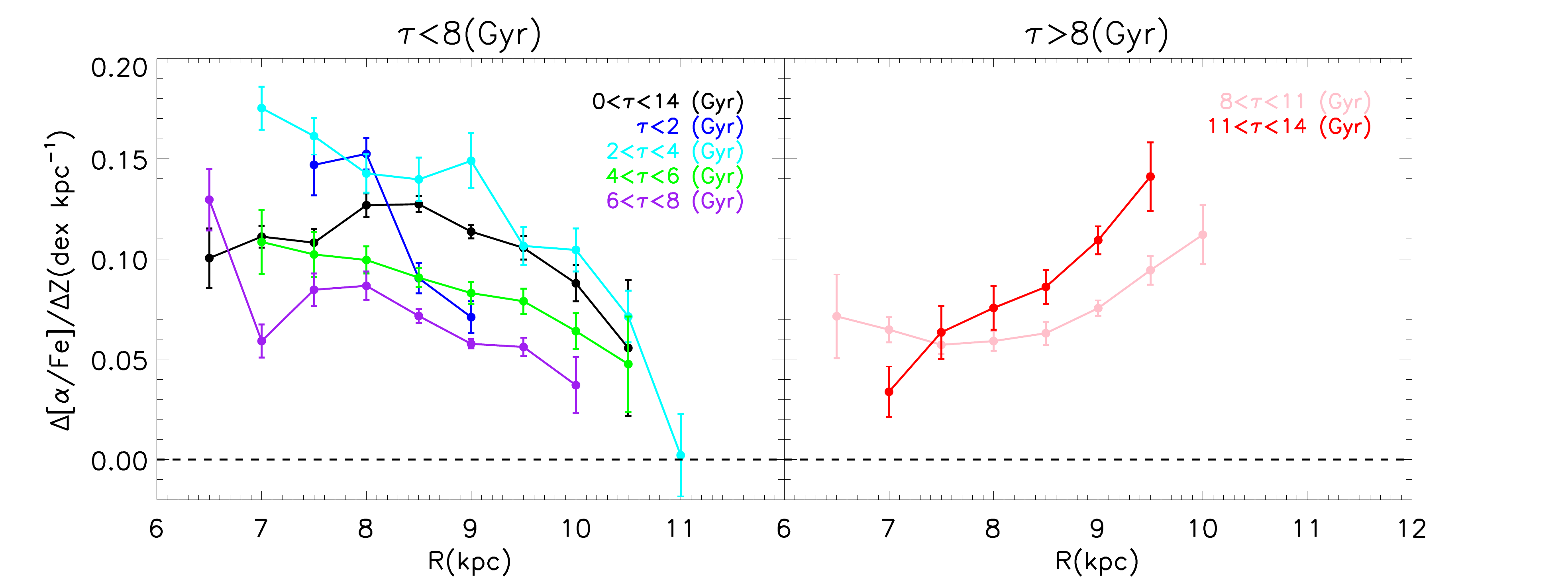}
\caption{Vertical [$\alpha$/Fe] gradients as a function of $R$ derived from mono-age stellar populations, presented with different colours, as marked in the plot. The horizontal dashed lines delineate a constant vertical [$\alpha$/Fe] gradient value of 0.0\,dex\,$\rm kpc^{-1}$. Vertical [$\alpha$/Fe] gradients flatten  as $R$ increases for populations of $\tau\,<\,8\,$Gyr (left panel), but steepen  as $R$ increases for populations of $\tau\,>\,8\,$Gyr (right panel). }

\label{afe_gradient_r}
\end{figure*}

 Fig.\ref{afe_gradient_r}  shows the  vertical [$\alpha$/Fe] gradients derived for the mono-age stellar populations as a function of $R$.  Almost all of the mono-age stellar populations have positive vertical  [$\alpha$/Fe] gradients at a given radial annulus.  
It also shows that the positive vertical [$\alpha$/Fe] gradients flatten with increasing $R$  for the stellar populations of $\tau < 8$ Gyr. The gradients steepen with increasing  $R$   for the stellar populations of $\tau > 8$ Gyr. 

 \subsection{Uncertainties of the gradients}

Throughout the paper, only the fitting errors are reported, and the main contribution are random errors of the stellar parameters ([Fe/H], $R$, $Z$ and age). In most cases, the fitting errors are very small ($<0.01\rm \,dex\,kpc^{-1}$) because of the large number of stars in our sample. We emphasize that those small values of gradient errors are not in conflict with the broad metallicity distribution of stars at a given $R$ or $Z$. This is because the latter is not only contributed by random errors of stellar parameters but also largely contributed by the intrinsic width of metallicity distribution. However, there are indeed other error (systematic) sources that are not considered in this work. For example, we derive the gradients with linear functions, but which may fail to describe the true radial or vertical trend of mean [Fe/H] or [$\alpha$/Fe] accurately. This is especially obvious in the case of the radial [$\alpha$/Fe] gradients (see Fig.\,\ref{afe_gradients}). In addition, contaminations from dwarf stars to our current MSTO-SG star sample are also expected to lead considerable systematic errors to the gradients of old stellar populations.

 \section{Profiles of the [Fe/H] and [$\alpha$/Fe] distributions}
 \subsection{Profiles of the [Fe/H] distributions}
 
In order to investigate the  profiles of the [Fe/H]  distributions,  we firstly map out the [Fe/H]  distributions of mono-age stellar populations in  each $R$ and  $|Z|$ bin.   In doing so, the individual  mono-age stellar populations  are divided into annuli of 1 kpc and 3kpc in the radial direction for $R < 12$ kpc and $R > 12$ kpc, respectively. In each annulus, we further divide the  mono-age stellar populations  into three bins of $|Z|$:   [0,0.3], [0.3,1.5] and [1.5,2] kpc, corresponding roughly to the   thin disc, the transition region between the  thin and thick disc and the thick disc, respectively.   Bins containing  less than 100 stars are discarded. 
Fig.\ref{mdf}  shows the normalised  [Fe/H]  distributions of the individual  mono-age stellar populations in  each $R$ and $|Z|$ bin.   In each bin, the histogram  [Fe/H] distribution is plotted with a bin size of  0.08\,dex, corresponding approximately to  half of the typical uncertainty of [Fe/H] measurements.   
The profiles of the [Fe/H] distributions of the different  mono-age stellar populations show significant variations with $R$ and $|Z|$, especially for the stellar populations of $\tau <$\,8\,Gyr.  For the stellar populations of $\tau <$\,8\,Gyr and $|Z| < 1.5 $ kpc, the [Fe/H] distributions vary from a negative skewed profile (having  relative long metal-poor tails) at the inner disc to a positive skewed profile (having relative long metal-rich tails) as $R$ at the outer disc.
For the stellar  populations  older than  8 Gyr at all $|Z|$ bins,  or the stellar populations  younger than 8\,Gyr  at high $|Z|$ bins, the distributions  show only weak variations with $R$. 
For the stellar population of $\tau >11$\,Gyr, our results also show an unexpectedly high fraction of metal-rich ${\rm [Fe/H]} >0$ stars, whose origins are not fully understood. A possibility is that they are  actually originated at the inner disc, and migrated to their current position due to kinematic processes. Contaminations of younger dwarf stars may have also contributed a significant part of them.   

\begin{figure*}
 \centering
 \includegraphics[width=7.0in]{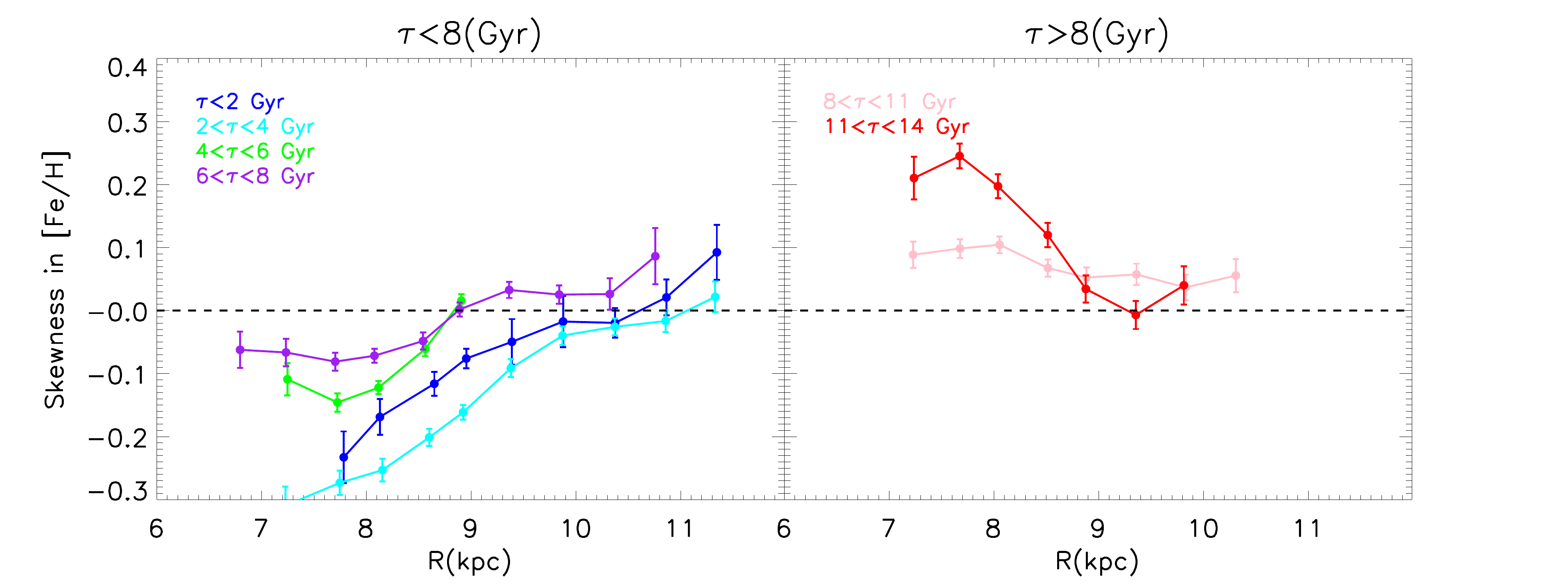}
 \caption{Skewness  as a function of $R$   for the  mono-age stellar populations (at $|Z| < 1.5$\,kpc),  presented with different colours, as marked in the plot. The skewness of stellar populations of $\tau\,<\,8\,$Gyr (left panel) increases as  $R$ increases, while that of stellar populations of $\tau\,>\,8\,$Gyr (right panel) decreases or is nearly invariant  as $R$ increases.  The results for stellar population of $\tau\,>\,11\,$Gyr  should be taken with caution due to the possible young dwarf star contamination.}
 
 \label{skewness_feh}
 \end{figure*}

 In order  to quantitatively describe   the differences  of the profiles of the  [Fe/H]  distributions of the the individual  mono-age stellar populations, we further derive the skewness of the distributions.  In doing so,  the individual  mono-age stellar populations  are divided into annuli of 1 kpc in the radial direction from 6 kpc to 13 kpc with a step of 0.5 kpc. The bins of $|Z|$ are the same as used for  Fig.\ref{mdf}.  Bins containing less than 600 stars are discarded to  ensure robust measurements of the skewness. Fig.\ref{skewness_feh} plots the radial variations of the skewness for mono-age stellar populations at $|Z|<1.5$\,kpc. The Figure shows a strong  radial trend of the skewness, and the trend varies with stellar age.   
 It shows that for populations of $ \tau < 8$ Gyr and $|Z|<1.5$\,kpc,  the skewness  increases with  increasing $R$ (switchs from negative to flat even positive),  and the radial trend becomes flatter  with increasing age. 
 For the old populations of $ \tau > 8$\,Gyr, the  skewness decreases with  increasing $R$. While the radial trend of the skewness of $8 < \tau < 11$\,Gyr is less significant.


  \cite{Hayden_2015} have  studied  the radial  variations of the profiles of the   [Fe/H] distributions (see their Fig.\,5) using  69,919 APOGEE red-giant stars. In their study they   divide the sample stars into two populations of  high and low [$\alpha$/Fe]  ratios, corresponding approximately  to the   old and young stellar populations, respectively.  
  Their results show that the [Fe/H] distributions of the low-[$\alpha$/Fe] population close to the Galactic plane ($|Z|<0.5$\,kpc) are negatively skewed in the inner disc and are positively skewed in the outer disc. Whereas the profiles of [Fe/H] distributions of the high-[$\alpha$/Fe] population, as well as those of the low-[$\alpha$/Fe] population at large height above the disc mid-plane, are almost symmetric and  invariant with $R$. These radial variations trends of the profiles of  [Fe/H] distributions  for low-[$\alpha$/Fe] and high-[$\alpha$/Fe] are generally in good agreement with our results for young ($<8$\,Gyr) and old ($>8$\,Gyr) stellar populations, respectively. Benefiting from the robust age estimates, we are able to determine the transition age  when  the profiles of the [Fe/H] distributions (more precisely, the skewness) change from the situation of  strong radial variations (at $\tau <$\,8\,Gyr) to a situation of  weak radial variations (at $\tau >$\,8\,Gyr). 

\begin{figure*}
\centering
\includegraphics[width=7.0in]{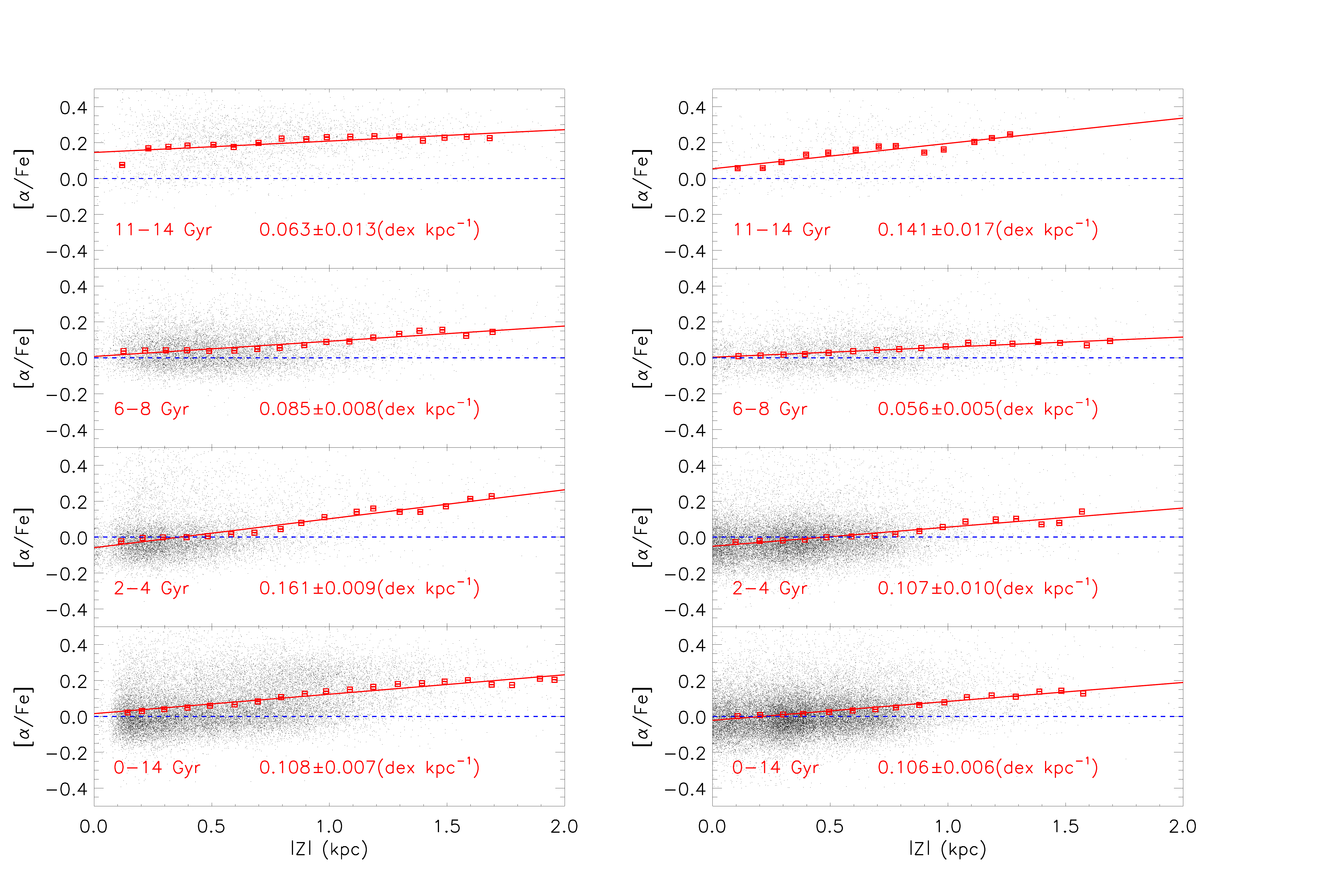}

\caption{Vertical  [$\alpha$/Fe] gradients derived from several individual  mono-age  stellar populations (as marked in each panel)  in  two radial bins: $7 < R < 8$ kpc (left panels) and $9 < R < 10$ kpc (right panels).  Red squares represent the CMD-weighted mean values of [$\alpha$/Fe] in the individual vertical bins.  The error bars of the mean values are also over plotted, but in almost all the cases they are smaller than the size of the symbols. Linear fits to red squares are over plotted by red lines. The blue  horizontal  dashed lines  delineate a constant [$\alpha$/Fe] value of $0.0$\,dex. The age range of each mono-age  stellar population,  the slope of the linear fit (vertical [$\alpha$/Fe] gradients) and the associated error are marked in each panel. The mean value of [$\alpha$/Fe]  as a function of $|Z|$ in each panel can be well fitted by a linear line.}

\label{afe_gradients_r}
\end{figure*}

 \subsection{Profiles of the [$\alpha$/Fe] distributions}
\begin{figure*}
 \centering
 \includegraphics[width=7.0in]{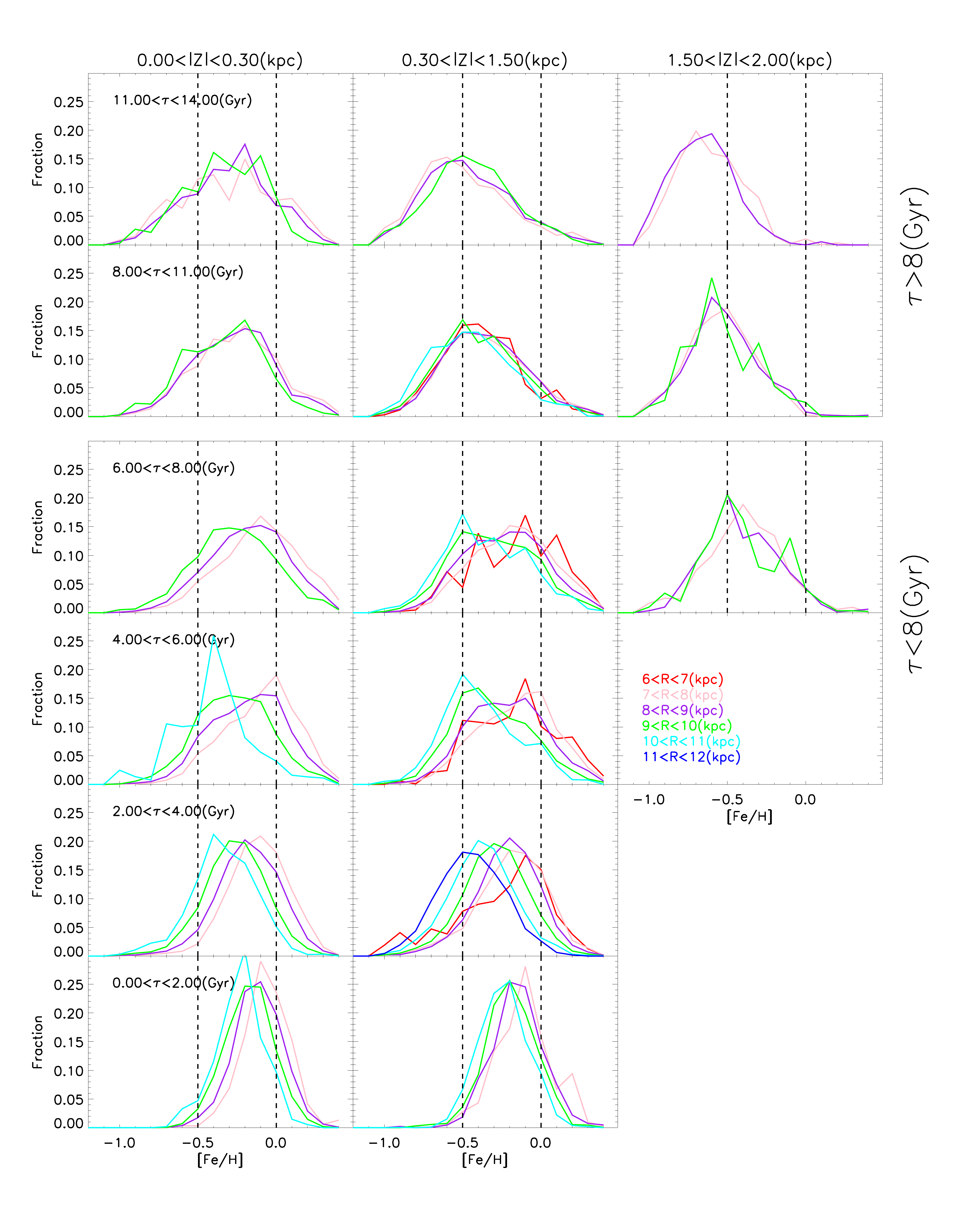}
 \caption{Distributions of [Fe/H]  in different $R$, $|Z|$ and $\tau$ bins. The $X$ axis show the values of [Fe/H],  $Y$ show the relative stellar numbers in  different [Fe/H] bins.  The left, middle and right panels show the results derived from stars at $|Z| < 0.3$ kpc, $0.3 < |Z| < 1.5$ kpc, 1.5\,$<$\,$|Z|$\,$<$\,2.0\,kpc, respectively. The age ranges of the  mono-age stellar populations are marked in each panel.  Results for  different radial bins are presented by lines of  different colours, as labeled in the lower right panel of the plot. The vertical dashed lines delineate  constant [Fe/H] values of $-0.5$ and 0.0\,dex. The profiles of [Fe/H] distributions show significantly radial variations  for young stellar populations ($\tau\,<\,8\,$Gyr), but no clear radial variations for old stellar populations ($\tau\,>\,8\,$Gyr).}
 
 \label{mdf}
 \end{figure*}

\begin{figure*}
 \centering
 \includegraphics[width=7.0in]{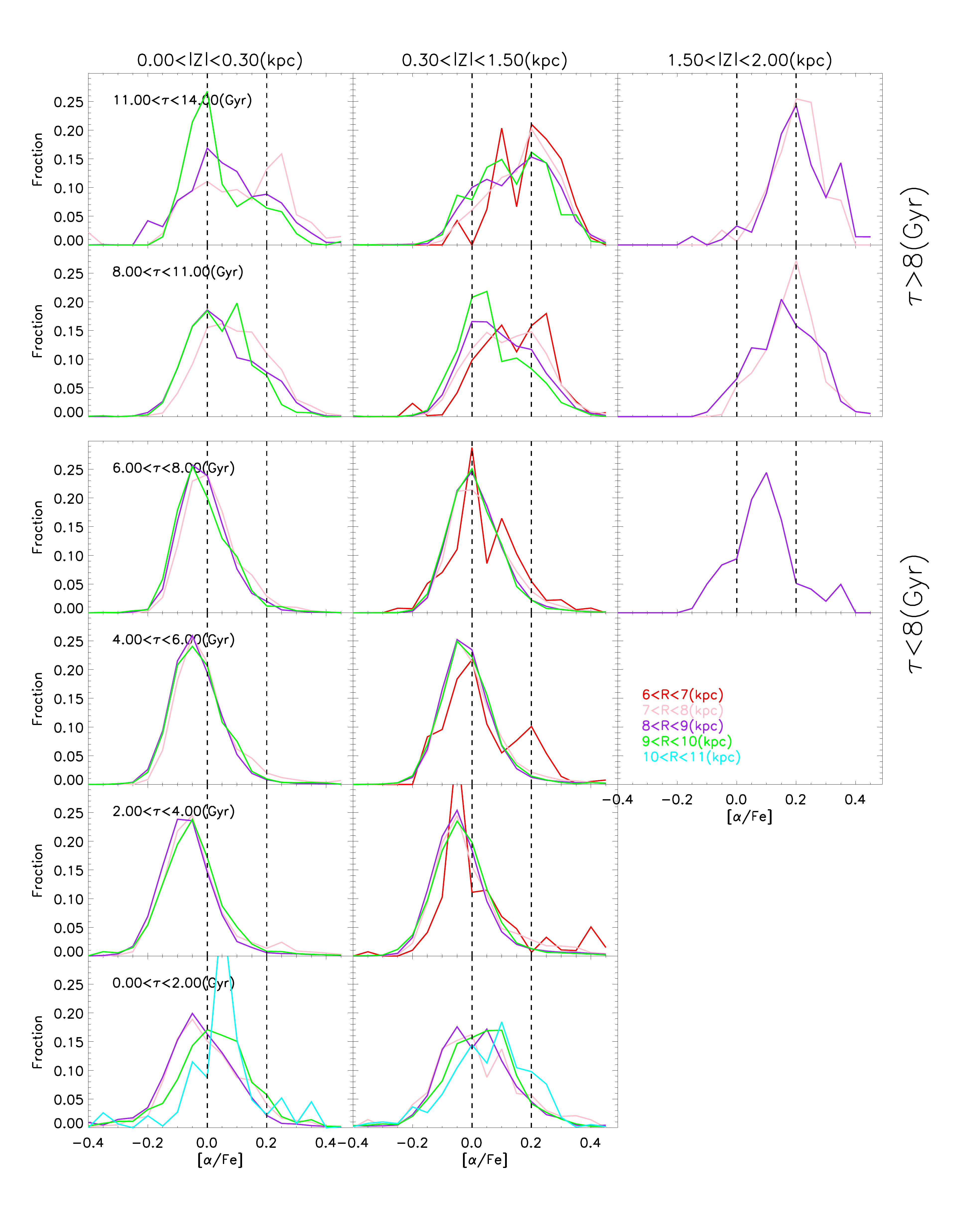}
 \caption{Same as Fig.\ref{mdf}, but for  [$\alpha$/Fe]. The vertical dashed lines delineate  constant [$\alpha$/Fe] values of $0.0$ and 0.2\,dex. The results for stellar population of $\tau\,<\,2\,$Gyr should be taken with caution  as about 10--20\,per\,cent  [$\alpha$/Fe] measurements of those stars  may be incorrect.}
 
 \label{afedf} 
 \end{figure*}

 To study the  profiles of the  [$\alpha$/Fe]  distributions of the mono-age stellar populations, the MSTO-SG-SNR50 sample is divided into  bins of  $R, |Z|$ and $\tau$. 
 The bins  used in terms of  $R, |Z|$ and $\tau$  are same as those adopted for  Fig.\ref{mdf}.  The normalised [$\alpha$/Fe] distributions are shown on Fig.\ref{afedf}. 
It shows that the profiles of the [$\alpha$/Fe] distributions vary with $R$ at all $|Z|$'s for  the populations of $\tau > 8$\,Gyr.  They  have  negatively skewed distributions in the inner disc and  positively skewed distributions in the outer disc.  For the stellar populations of   $2 < \tau < 8$ Gyr, the profiles of  the [$\alpha$/Fe] distributions  vary only  weakly with $R$ and $|Z|$.  The results also show that the stellar population of   $4 < \tau < 6$ Gyr at $R < 7$\,kpc contain much more $\alpha$-rich stars than at $R > 7$\,kpc. Considering the small number of stars in this bin, we presume this result is not significant.   The profiles of   [$\alpha$/Fe] distributions of stellar populations of $\tau\,<\,2\,$Gyr show variations with $R$. However, we note that the stellar populations of $\tau\,<\,2\,$Gyr  contains an unexpectedly high fraction of high [$\alpha$/Fe] stars, and this may be the consequence of incorrect [$\alpha$/Fe] estimates of some of those stars.
 As  discussed by \cite{Xiang_msto},  those objects  are likely spectroscopic binaries of composite spectra. The  [$\alpha$/Fe] values of those stars have been  incorrectly estimated. The fraction of youngest stars with incorrect  values of [$\alpha$/Fe]  is about 10--20 per cent.  Thus, the profiles of  [$\alpha$/Fe] distributions of youngest stars should be taken with caution.

\begin{figure}
 \centering
 \includegraphics[width=3.5in]{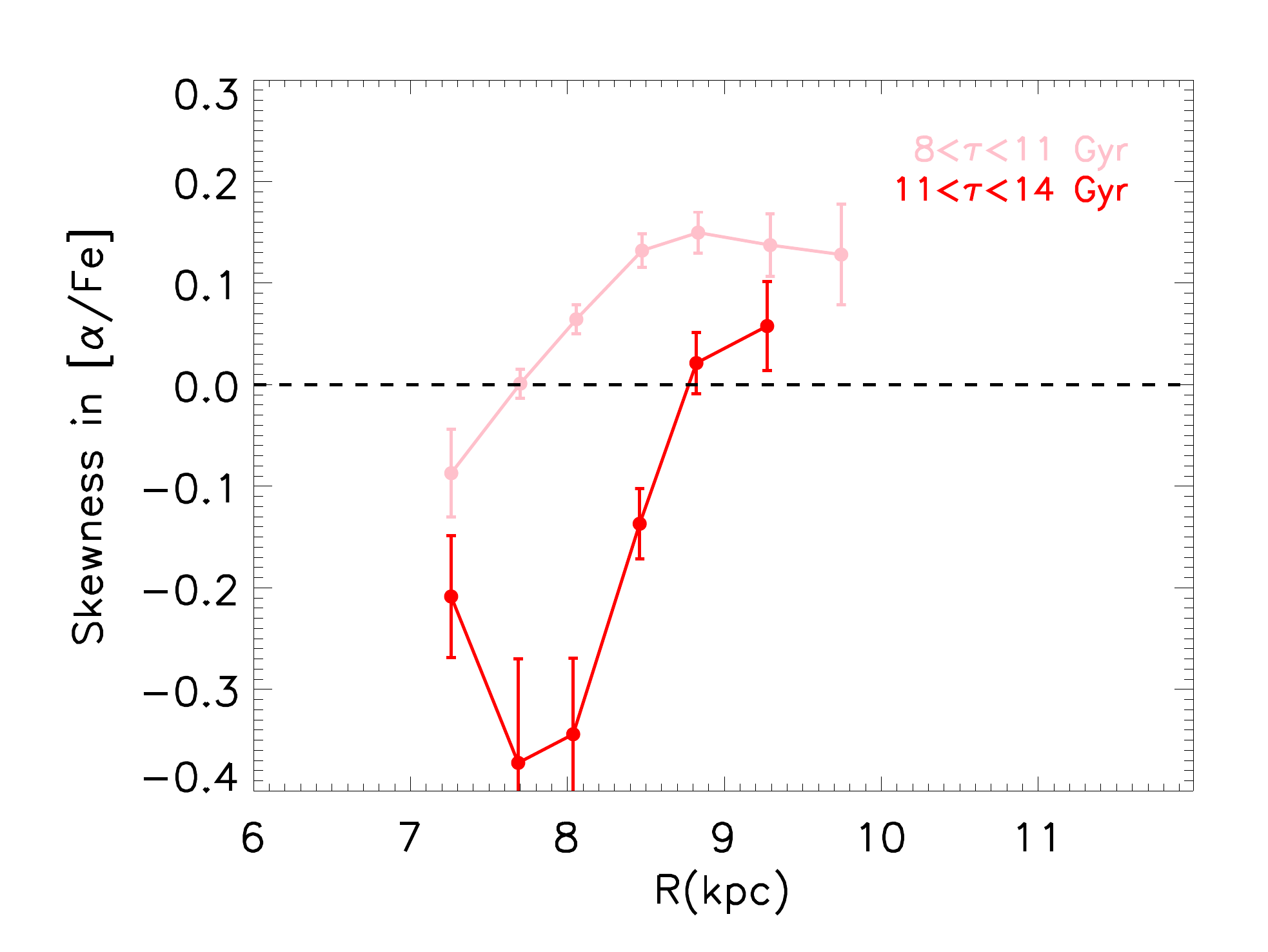}
 \caption{Same as Fig.\ref{skewness_feh}, but for  [$\alpha$/Fe]. The skewness of [$\alpha$/Fe] distributions of old stellar populations ($\tau\,>\,8$\,Gyr) increase as $R$ increases.} 
 
 \label{skewness_alpha}
 \end{figure}

Values of  skewness of the  [$\alpha$/Fe] distributions are also estimated for the mono-age  stellar populations, as shown in Fig.\ref{skewness_alpha}. 
As discussed in the above paragraph, we  find that the profiles of [$\alpha$/Fe] distributions of  $\tau <$ 2.0 Gyr are likely to be unrealistic and that of $2 < \tau < 8$ Gyr vary weakly with $R$. Thus, values of skewness of  the [$\alpha$/Fe] distributions of  the young stellar populations  are not presented here. 
  Fig.\ref{skewness_alpha}  shows only the radial variations of the skewness of the [$\alpha$/Fe] distributions for  the stellar populations of $\tau >$ 8 Gyr in different $|Z|$ bins. 
It shows that the skewness of the  stellar populations of $\tau >$ 8 Gyr  increase with  increasing  $R$ at $|Z| < 1.5$ kpc.  

\section{Impacts of radial migration on the metallicity distribution function}

\begin{figure*}
 \centering
 \includegraphics[width=7.0in]{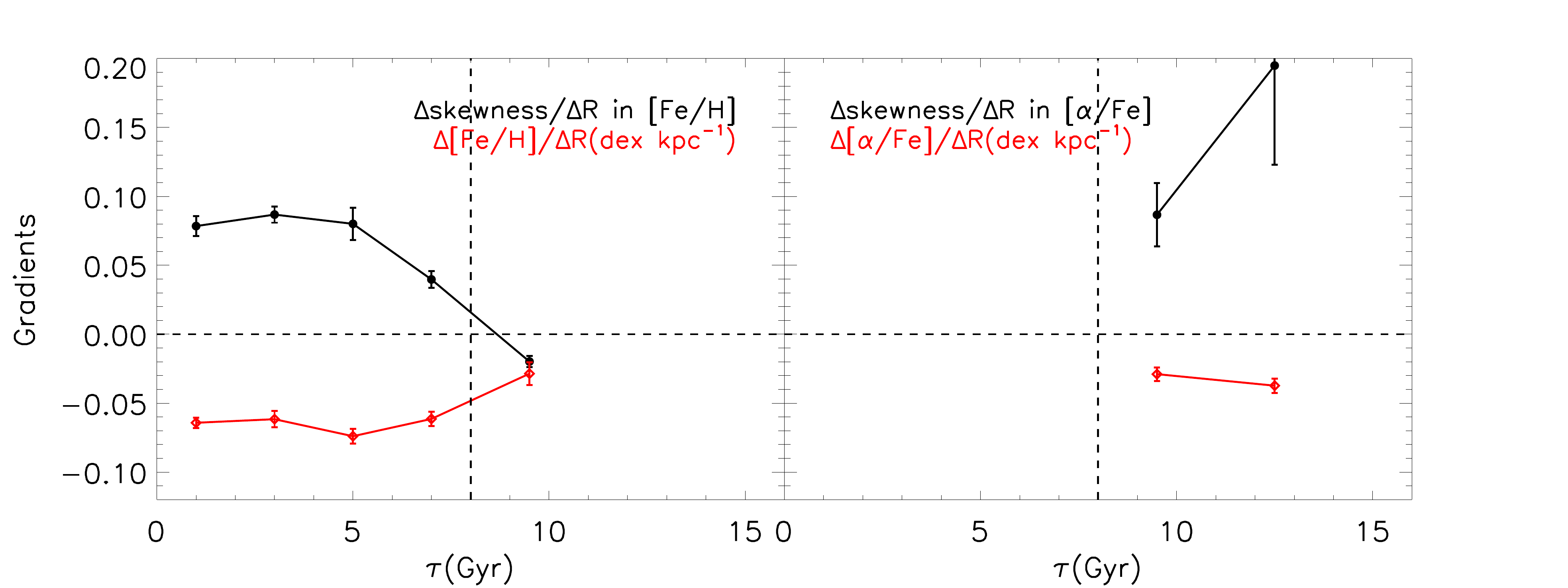}
 \caption{The radial gradients of  skewness (black dots and lines) and mean values (red dots and lines) of [Fe/H] (left panel) and [$\alpha$/Fe] (right panel)  as a function of age for stellar populations at $|Z|\,<\,1.5\,$kpc. The horizontal lines delineate a constant gradient value of 0.0. The vertical lines delineate a constant age  of 8\,Gyr. The variations of the radial skewness gradients with age are opposite to the  variations of the radial gradients of the mean values. }
 
 \label{skewness_gradient}
 \end{figure*}

The  radial dependence  of the profiles of the [Fe/H] distributions  has been modelled by  \cite{Schonrich} and \cite{Toyouchi2018} using a chemical evolution model including the radial migration processes.  
The radial migration processes carry stars from the  inner (metal-rich) or  outer (metal-poor) disc region  to their current spatial position ($R$, $Z$), and consequently  change the profiles of the [Fe/H] distributions owing to the existence of  radial metallicity gradients of the Galactic disc.  
Because the Galactic disc exhibits negative  radial [Fe/H] gradients, the inwards migrators generally contribute the metal-poor  tail  of the [Fe/H]  distribution, while the outwards migrators contribute mainly the metal-rich  tail. 
 The process will produce a relatively negative  skewed [Fe/H]   distribution in the inner disc and a relatively positive  skewed [Fe/H]  distribution in the outer disc, and consequently produce a positive radial gradient of skewness. 
 That is the radial gradients of skewness and mean values and their evolution with age should be opposite in sign.
Fig.\,\ref{skewness_gradient} shows the radial gradients of  skewness and mean values of [Fe/H] and [$\alpha$/Fe]  as a function of age for stellar populations at $|Z|\,<\,1.5$\,kpc.  It shows that the radial gradients of skewness and mean values and their evolution with age are indeed  opposite in sign, and the stronger the abundance gradients, the stronger the gradients in skewness.
The radial [Fe/H] gradients of stellar populations of $\tau < 8$ Gyr   at $|Z|\,<\,1.5$\,kpc are negative, while the radial gradients of skewness in [Fe/H] are positive. For stellar populations of $\tau > 8$ Gyr   at $|Z|\,<\,1.5$\,kpc, both the radial gradients of skewness and of mean values of [Fe/H]  are  nearly flat. 
This suggests that the radial migration  is clearly in operation and has affected the profiles of the [Fe/H] distributions of stellar populations with significant radial [Fe/H] gradients.  The absence of the radial variations   of the profiles of  the [Fe/H] distributions  for    stellar populations of   $\tau > 8$ Gyr is likely caused by  the absence of any significant  radial [Fe/H] gradients of those stellar populations. 
Fig.\,\ref{skewness_gradient} also shows that the radial [$\alpha$/Fe]  gradients of stellar populations of $\tau > 8$ Gyr are negative, while the radial skewness gradients in [$\alpha$/Fe] are positive.  The radial variations of profiles of [$\alpha$/Fe] for stellar populations of $\tau\,>\,8\,$Gyr are likely  the combined consequence  of the presence  of negative radial  [$\alpha$/Fe] gradients and the radial migration processes. 

In short,  negative radial  [Fe/H] or [$\alpha$/Fe] gradients, combined with the radial migration processes, which change the motion radii of the stars,  are responsible for the observed radial variations  of the profiles of the [Fe/H] and [$\alpha$/Fe]  distributions for young and old populations, respectively. 

Generally, there are two types of radial migrations, the blurring and churning \citep[e.g. ][]{Schonrich}. In the churning process, a star changes its orbit radii due to a  gain/loss of angular momentum. In the blurring process, the angular momentum of a star is conserved and its epicycle amplitude changes.    In the remaining parts of this Section, we will further discuss the effects of the blurring process on the metallicity distributions of the mono-age stellar populations. First, we integrate the orbit of each star, assuming  a Galactic  gravitational potential  given by \cite{Gardner2010}, for a  long enough time and derive the average $R$ and $|Z|$, i.e.,  $<R>$ and $<|Z|>$.   We then examine the  distributions   as a function of  the guiding centre radius  ($R_{\rm g}\,=\,<R>$) and $|Z|$,  rather than $R$ and $|Z|$.   The results seem   to rule out any significant  effects of blurring in play. 

\begin{figure*}
\centering
\includegraphics[width=7.0in]{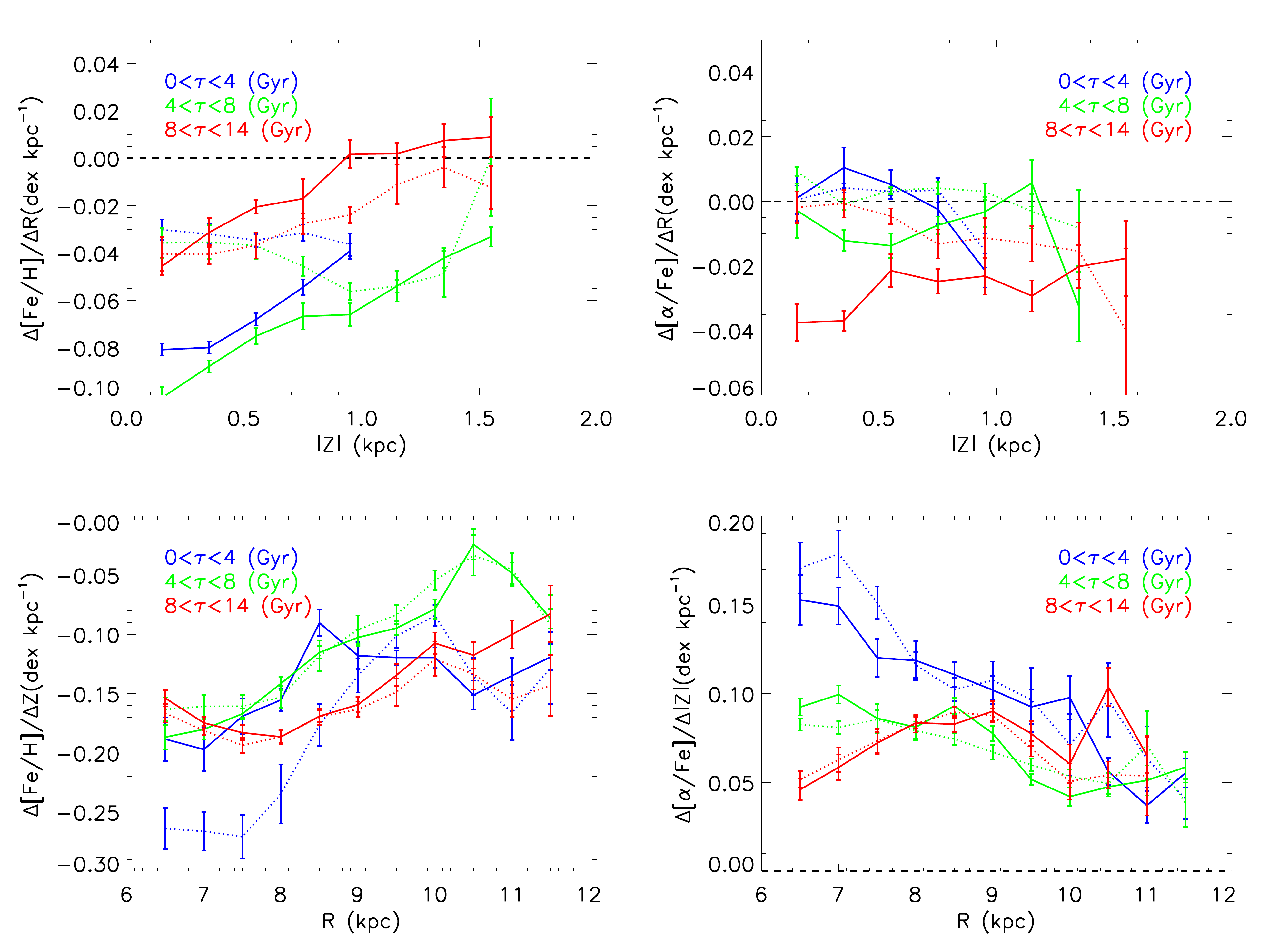}

\caption{Radial (top panels) and vertical (bottom panels) [Fe/H] (left panels) and [$\alpha$/Fe]  (right panels) gradients. Solid lines show  gradients estimated in the $R$--$|Z|$ plane, while  dotted lines show those estimated in the  $R_{\rm g}$--${|Z|}$ plane. Different colors show the results of the different mono-age stellar populations, as marked in the panels. The horizontal  lines delineate a constant gradient value of  0.0\,dex\,$\rm kpc^{-1}$. } 
\label{gradient_r_rg}

\end{figure*}

Fig.\ref{gradient_r_rg}  shows the radial and vertical [Fe/H] and [$\alpha$/Fe] gradients estimated in the  $R$--$|Z|$  and $R_{\rm g}$--${|Z|}$ planes for the  different mono-age stellar populations.  
For stellar populations of $\tau > 8$\,Gyr, the radial [Fe/H] and [$\alpha$/Fe] gradients estimated  in the $R$--$|Z|$ plane are flatter and steeper than those derived in the  $R_{\rm g}$--${|Z|}$ plane, respectively. In the regions of low $|Z|$, the radial [Fe/H] gradients derived in $R$ and $|Z|$ plane are steeper  than   those deduced in the $R_{\rm g}$--${|Z|}$ plane for the stellar populations of $\tau < 8$\,Gyr.   The radial [$\alpha$/Fe] gradients for the stellar populations of $4 < \tau < 8$ Gyr derived in the $R$--$|Z|$ plane are steeper than those derived in the  $R_{\rm g}$--${|Z|}$ plane. 
The results  suggest that the blurring process may steepen  the radial [Fe/H]  gradients for populations of $\tau < 8$ Gyr,  and the radial [$\alpha$/Fe]  gradients for all stellar populations,   while its effects on the radial [Fe/H]  gradients are likely to be small or negligible for populations of $\tau > 8$ Gyr. The results are opposite to the predictions of chemical evolution models \citep[e.g. ][]{Schonrich} that  suggest that the radial migration  will lead to the flattening of the radial metallicity gradients.  While the chemical evolution models always assume a smooth radial stellar number (mass) distribution,  the real disc shows complicated structure, and this may partly account for the difference. It is noted that \cite{Schonrich} considered  both two types of radial migration processes, which may also produce the difference.  The vertical [Fe/H] and  [$\alpha$/Fe] gradients derived in the  $R$--$|Z|$ plane and  those derived in the $R_{\rm g}$--${|Z|}$ plane are quite identical to each other with an exception that the vertical [Fe/H] gradient at $R$--$|Z|$ plane is  flatter than that at $R_{\rm g}$--${|Z|}$ plane for population of $\tau < 4$\,Gyr. 

\begin{figure*}
\centering
\includegraphics[width=7.0in]{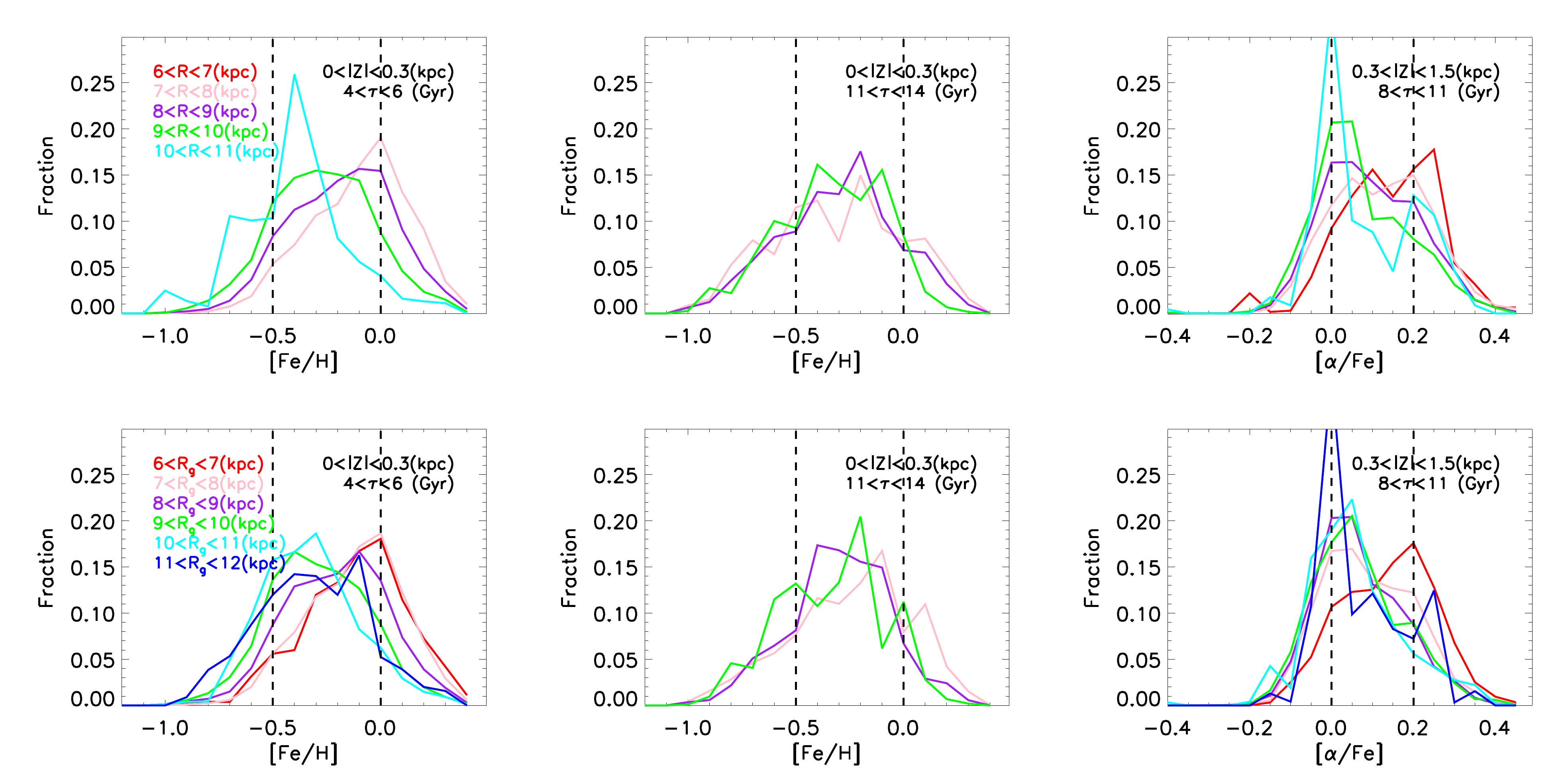}

\caption{[Fe/H] (first and second columns) and [$\alpha$/Fe] (third column) distributions. The top panels show the distributions derived in the $R$--$|Z|$ plane, while those in the  bottom show the  results derived in the $R_{\rm g}$--$|Z|$ plane. Panels in the first column show the [Fe/H] distributions derived for the stellar populations of $4 < \tau < 6 $ Gyr close to the Galactic plane. The panels in the second column show the [Fe/H] distributions derived for the stellar populations of  $11 < \tau < 14 $ Gyr close to the Galactic plane. The panels in the third column show the [$\alpha$/Fe] distributions for the stellar populations of $8 < \tau < 11 $ Gyr  of height $0.3 < |Z| < 1.5$ kpc. The vertical lines delineate constant [Fe/H] values of $-0.5$ and 0.0\,dex, and constant [$\alpha$/Fe] values of 0.0 and 0.2\,dex. } 
\label{mdf_r_rg}

\end{figure*}

 As discussed in Section 4, the skewness of the [Fe/H]  and [$\alpha$/Fe]  distributions is a function of $R$ for the young and old populations, respectively.  In order to check whether the blurring is the main process producing the radial trends of [Fe/H]  and [$\alpha$/Fe]  distributions, we plot  the profiles of the [Fe/H] and [$\alpha$/Fe] distributions at different $R$ and $R_{\rm g}$ bins for  the young and old  populations at relatively low $|Z|$ regions, respectively, in Fig.\ref{mdf_r_rg}. 
In Fig.\ref{mdf_r_rg}, we plot the distributions of [Fe/H] of stellar populations of $4\,<\,\tau\,<\,6\,$Gyr and $11\,<\,\tau\,<\,14\,$Gyr in order to check whether the blurring effect affects the [Fe/H] distributions of different mono-age populations differently.  For the distributions of [$\alpha$/Fe], we only show the results of stellar populations of $8\,<\,\tau\,<\,11\,$Gyr at $0.3\,<\,|Z|\,<\,1.5$\,kpc in which there are enough number of stars. Fig.\ref{mdf_r_rg} shows that the profiles of the [Fe/H] and [$\alpha$/Fe]
distributions also vary with $R_{\rm g}$  for the young and old populations, respectively. This  suggests that  the radial variations of the skewness is not a consequence  of the blurring process.  \cite{Hayden_2015}   argue that a chemical evolution model considering the churning  rather than the blurring process   can reproduce the radial variations  of the observed profiles  of the [Fe/H] distributions. Alternatively, the effects of blurring must be much  smaller than those of churnning.  Their conclusions are  consistent with our results. 

Fig.\ref{mdf_r_rg} also shows that the dispersions of the metallicity distributions derived in the $R$--$|Z|$ plane are larger than those derived in the $R_{\rm g}$--$|Z|$ plane. This suggests that the blurring process may have increased the dispersions of the metallicity distributions for both the young and old stellar populations.   Comparing the top and bottom panels in the third column of the Figure, one  finds that there are more high-$\alpha$ stars in the $7 < R < 8$ kpc and $0.3 < |Z| < 1.5$\,kpc bin than  in the $7 < R_{\rm g} < 8$ kpc and $0.3 < |Z| < 1.5$ kpc bin. By examining  the $R_{\rm g}$ distribution of stars in $7 < R < 8$ kpc, we find that there are much more stars of $R_{\rm g} < 7$ kpc than stars of $R_{\rm g} > 8$ kpc.  Clearly, those high-$\alpha$ stars originate from the inner disc and move to their current positions  through the blurring process.

  \section{Constraining  the formation and evolution of the  disc(s)}
  
  \subsection{The formation scenario of the disc at an early epoch}
The radial and vertical [Fe/H]  and [$\alpha$/Fe] gradients of the oldest stellar populations provide strong constrains on the  formation scenario of the disc at early epoch. 
The oldest stellar  populations ($\tau >$ 11 Gyr) show nearly flat or marginally positive radial [Fe/H] gradients. 
It suggests that the oldest stellar populations formed from gas that was originally  well-mixed in the radial direction in  a violent, fast formation process.   A fast, highly-turbulent gas-rich merger formation scenario
as proposed by \cite{Brook2004,Brook2005}  suggests that the thick disc formed so fast (with a  timescale shorter than the life time of Type Ia supernovae (SNe Ia), which produce most of the iron elements)  at an early epoch  that there should be no [Fe/H] gradients in the radial  direction, consistent with our results of the oldest stellar populations.

On the other hand, we find that there are significant  negative vertical [Fe/H] gradients and negative radial [$\alpha$/Fe] gradients for the oldest stellar populations.  This suggests that the formation of the oldest disc (meaning here the disc older than 11 Gyr) can  not  be too  fast as it needs enough time to develop  vertical [Fe/H] gradients and negative radial [$\alpha$/Fe] gradients.  The timescale of disc formation should be longer than the life time of Type II supernovae (SNe II).  SNe II produce most of the $\alpha$ elements.   Before 11 Gyr, several generations of stars  have evolved and changed the  [$\alpha$/Fe] of the disc.  Higher star formation rates (SFRs) (discussed in the Section\,6.2) in the inner disc produce more exploring of SNe II in the inner disc and  the negative radial [$\alpha$/Fe] gradient presented in the current work.


In conclusion, the disc formed from an originally well-mixed  gas  in  a violent, fast formation process at early epoch. The timescale of the process is shorter than the life time of SNe\,Ia, but longer than the life time of SNe\,II. In the evolution of the oldest disc, the vertical heating play an important role. 

 \subsection{The ``inside-out'' and ``upside-down'' disc formation scenario}

Our data  shows negative radial [Fe/H] gradients for  the stellar populations younger than 11 Gyr and negative radial [$\alpha$/Fe] gradients for the oldest stellar populations , which is  a consequence of  the ``inside-out'' disc formation scenario \citep{Matteucci1989,schonrich2016}.  In this scenario, SFRs  in the inner disc are higher than  in the outer disc, yielding  more metals  in the inner disc than in the outer disc and the negative radial [Fe/H] gradients.  As discussed in the  Section\,6.1, the ``inside-out'' disc formation scenario  is also responsible for the negative radial [$\alpha$/Fe] gradients for the oldest stellar populations.  At early epoch, as $R$ decreases, the SFRs increase,  yielding  more exploding of SNe II, more $\alpha$-elements and higher [$\alpha$/Fe] in the inner disc.

All the  mono-age stellar  populations show  negative vertical [Fe/H] gradients. 
This is  consistent with the theoretical predictions of \cite{Kawata}, who consider among others also mono-age stellar populations formed in a flaring, star-forming disc. In this scenario,   stars in the outer regions can  reach to  higher vertical heights than those stars in the inner regions because of their higher vertical energies, which producing a flaring disc.  Radial mixing then yield negative vertical [Fe/H] gradients  \citep{Kawata, Ciuca2018}. 
 If the radial migration process is slow,  old stars have more  time to migrate to  outer disc,  producing a steeper  (negative) vertical [Fe/H] gradients for the old stellar populations. 
 Again this is consistent with our finding. 
On the other hand,  if  this scenario does work, then one would expect negative vertical  [$\alpha$/Fe] gradients,  not  consistent with our finding  that all the mono-age  stellar  populations   exhibit  positive vertical  [$\alpha$/Fe] gradients. 
 Our results thus prefers  the ``upside-down'' disc formation scenario.  This  scenario suggests that the  height of the star-forming gas
layer contracts as the stellar mass of the disc grows \citep{Bird2013,Bournaud}.  And as the  metal-rich and [$\alpha$/Fe]-poor gas settles down to the Galactic plane, it produces  negative  vertical [Fe/H] gradients and positive vertical [$\alpha$/Fe] gradients.

 \subsection{The superposition of two discs} 

For the  low-$|Z|$  regions, our current study finds that the  radial  [$\alpha$/Fe] gradients   steepen with increasing age, while for the  high-$|Z|$ regions, the gradients flatten with  increasing age.  
The results are  the consequence of  superposition of two discs, a low-$\alpha$, ``thin'' disc of a  large scale length, and a  high-$\alpha$, ``thick'' disc of a  small scale length.  The relative fraction of the ``thin'' and the ``thick'' disc are different for different mono-age stellar populations, younger stellar populations contain more low-$\alpha$, ``thin'' disc stars. 
In the low-$|Z|$ regions,  low-$\alpha$ ``thin'' disc stars dominate both  the inner and outer discs (see Fig.\ref{afe_gradients_r}) for young stellar populations, leading to  relatively flat gradients. Populations of   older ages contain  more $\alpha$-rich ``thick'' disc stars in the inner disc than in the outer disc (see Fig.\ref{afe_gradients_r}), yielding  relatively steeper (negative) gradients. 
In the high-$|Z|$ regions, ``thick''  disc, $\alpha$-rich stars dominate  in the inner disc (see Fig.\ref{afe_gradients_r}) for all the mono-age stellar  populations.  
It seems that the  $\alpha$-rich disc contains stars as young as $2-4$ Gyr.  Those young $\alpha$-rich stars are either a consequence of a sustained star formation process of the thick disc, or intrinsically old stars whose ages have been incorrectly  underestimated--some of them may well be  blue straggles \citep{Xiang_msto}.  
In the outer disc,  young stars  are dominated by ``thin'' disc, $\alpha$-poor  stars (see Fig.\ref{afe_gradients_r}). This produces  relatively steeper negative radial [$\alpha$/Fe] gradients for young stars. Likewise,  old stars in the outer disc are dominated by ``thick'' disc,  $\alpha$-rich stars, leading to  relatively flat  negative radial [$\alpha$/Fe] gradients for old stars.  The superposition of two discs can also naturally explain the finding that the positive vertical [$\alpha$/Fe] gradients flatten with increasing $R$  for the stellar populations of $\tau < 8$ Gyr, while steepen with increasing  $R$ for the stellar populations of $\tau > 8$ Gyr.

\section{SUMMARY}
In this work, two samples, a  sample   consisting of 453,188 LAMOST MSTO-SG stars of spectral SNRs\,$>$\,20 (MSTO-SG-SNR20) and another consisting of 291,126 similar stars of  spectral SNRs\,$>$ 50 (MSTO-SG-SNR50), are built to study the 
[Fe/H] and [$\alpha$/Fe] distributions of mono-age stellar populations, respectively. Here, we have carried out for the first time a systematic  study of  the profiles of [Fe/H] distributions, the radial and vertical [$\alpha$/Fe] gradients and the profiles of [$\alpha$/Fe] distributions of mono-age stellar populations.  For the MSTO-SG-SNR20 sample, estimates of  effective temperature,  surface gravity, metallicity, radial velocity  and absolute magnitudes are estimated to be accurate to  150\,K, 0.2 dex, 0.15 dex, 5\,km\,$\rm s^{-1}$ and 0.5--0.6 mag, respectively.  Estimates of  [$\alpha$/Fe] are estimated to be accurate to  0.05 dex for MSTO-SG-SNR50 sample stars.   For the two MSTO-SG samples, age and distance errors are about 30  and 17 per\,cent, respectively.     The main results  presented in this work are  the following:

(a) The main results on the radial and vertical [Fe/H] and [$\alpha$/Fe]  gradients:
 \begin{itemize}
 \item Radial [Fe/H] gradients derived from  stars of $\tau <$\,11\,Gyr are negative, and vary with  $|Z|$ and $\tau$. The  gradients flatten with increasing $|Z|$. Radial [Fe/H] gradients derived from stars of $\tau >$\,11\,Gyr are  $\sim$\,$-$0.00$\pm$0.03\,dex\,$\rm kpc^{-1}$,  and vary little  with $|Z|$.  After being nearly flat at the earliest epoch of the disc formation,  the radial [Fe/H] gradients  steepen with decreasing age, reaching  maxima between $4 < \tau < 6$ Gyr, and then flatten again. 
\item There are negative vertical [Fe/H] gradients, that vary with $R$ and $\tau$. The  gradients flatten with increasing $R$, and steepen  with increasing age at $R > 8$ kpc. 
\item Radial [$\alpha$/Fe] gradients derived for the  stellar populations of $\tau < 4$ Gyr decrease with increasing $|Z|$.  The  gradients are nearly flat for the stellar populations of $4 < \tau < 6$ Gyr  and invariant with  $|Z|$. There are negative gradients for the stellar  populations of $\tau > 6$ Gyr. The   gradients is a function of age. They steepen with increasing age in the low-$|Z|$ regions and flatten with increasing age in the  high-$|Z|$ regions. 
\item There are positive vertical [$\alpha$/Fe] gradients that vary with $R$ and $\tau$. The gradients  flatten  with increasing $R$   for the  stellar populations of  $\tau < 8$ Gyr and    steepen with increasing $R$ for the stellar populations of $\tau > 8$ Gyr. 
\end{itemize}

 (b)The main results of profiles on the [Fe/H] and [$\alpha$/Fe] distributions: 

 \begin{itemize}
\item The profiles of the [Fe/H] distributions  vary with $R$  for  the young stellar populations of $ \tau < 8$ Gyr  close to the Galactic plane ($|Z| < 1.5$ kpc). They  show   negatively skewed  distributions in  the inner disc and  positively skewed distributions in the outer disc. 
The profiles of the distributions  of the young stellar populations in the high-$|Z|$ ($|Z|$\,$>$\,1.5\,kpc) regions show only marginal variations  with $R$.
The profiles  are nearly the same  in all $R$ bins for  the stellar  populations of $\tau > 8$ Gyr in all $|Z|$ regions. 
\item The profiles  of the  [$\alpha$/Fe] distributions change with $R$ for the old stellar populations of $\tau > 8$ Gyr in all $|Z|$ regions. Old stellar populations have  negatively skewed  distributions in the inner disc and  positively  skewed   distributions in the outer disc.   The profiles  show weak radial variations for  populations of $\tau < 8$ Gyr. 
 \end{itemize}

Radial  and vertical [Fe/H] gradients and their  variations with radius and height deduced here for mono-age populations are quite similar with those obtained by  \cite{Xiangmdf}. 
For the stellar populations of $ \tau < 8$ Gyr, the observed radial variations of the profiles of the  [Fe/H] distributions  also match very well  with those  derived from  the $\alpha$-poor  stellar   populations   by \cite{Hayden_2015}, who  suggest that  radial migration (especially the churning process)  is an important  process in the Galactic disc formation and evolution. The radial variations of the profiles of  the [Fe/H] distributions  derived here for the   stellar populations of $\tau > 8$ Gyr also  match very well with those derived from the  $\alpha$-rich stellar populations by   \cite{Hayden_2015}.
 
By examining  the metallicity distributions in the $R$, $|Z|$ and $\tau$ space, as well as in the  $R_{\rm g}$, ${|Z|}$ and $\tau$ space, we find that the churning process rather than the blurring process is the main reason responsible for the observed radial variations of the profiles of the metallicity distributions, consistent with the results of \cite{Hayden_2015}.  The blurring process may have   increased the widths  of the profiles of the metallicity distributions. 

Benefiting from  the wide age and large volume coverage of our MSTO-SG samples, we have mapped the metallicity distributions   in different $R$, $|Z|$ and $\tau$  bins. The results  provide strong constraints of  formation and evolution of the Galactic disc(s).  Our measured   radial [Fe/H]  and [$\alpha$/Fe] gradients support the   ``inside-out''   disc formation scenario, while  the vertical [Fe/H]   and  [$\alpha$/Fe] gradients  are consistent with  the ``upside-down'' disc  formation scenario.  Variations with age of the radial  [$\alpha$/Fe] gradients suggest that the Galactic disc is a superposition of two discs, a  low-$\alpha$, ``thin'' disc of a large scale length,  and a  high-$\alpha$, ``thick'' disc of a small scale length.
The measured values of  [$\alpha$/Fe]  provide strong constraints of the SNe\,II/Ia ratio. The clearly observed  negative radial  [$\alpha$/Fe] gradients derived from the  populations of $\tau > 8$ Gyr suggest that SNe II/Ia ratio is larger in the outer disc than that in the inner disc at early epoch of the disc formation.

\section{ACKNOWLEDGEMENTS}
We appreciate the helpful comments of the anonymous referee. 
This work is supported by National Key Basic Research Program of China
2014CB845700 , the NSFC grant 11703035 and the Joint Funds of the National Natural Science
Foundation of China (Grant No. U1531244 and U1331120. Guoshoujing Telescope (the Large Sky Area Multi-Object Fiber Spectroscopic
Telescope LAMOST) is a National Major Scientific Project built by the Chinese Academy of Sciences.
Funding for the project has been provided by the National Development and Reform Commission.
LAMOST is operated and managed by the National Astronomical Observatories, Chinese Academy of
Sciences. This work is also supported by National Natural Science Foundation of China (Grant No.
11473001). 
\bibliographystyle{mn2e}

\bibliography{mdf}

\end{document}